\documentclass[showpacs,showkeys,aps,prd,amsfonts,eqsecnum,nofootinbib,
floatfix,natbib]{revtex4-1}
\pdfoutput=1
\usepackage[T1]{fontenc}
\usepackage{amsmath,amssymb}
\usepackage{graphicx}
\usepackage{xcolor}
\usepackage{cancel}

\newcommand{\lsim}{\mbox{\raisebox{-.6ex}{~$\stackrel{<}{\sim}$~}}}

\newcommand{\met}{\ensuremath{\not\!\!E_T}\xspace}
\newcommand{\be}{\begin{equation}}
\newcommand{\ee}{\end{equation}}
\newcommand{\bea}{\begin{eqnarray}}
\newcommand{\eea}{\end{eqnarray}}

\newcommand{\newc}{\newcommand}
\newc{\bi}{\begin{itemize}}
\newc{\ei}{\end{itemize}}

\newc{\ra}{\rightarrow}
\newc{\sq}   {\mbox{$\wt{q}$}}
\newc{\msq}  {\mbox{$m_{\sq}$}}
\newc{\gl}   {\mbox{$\wt{g}$}}
\newc{\mgl}  {\mbox{$m_{\gl}$}}
\def \met  {\mbox{${E\!\!\!\!/_T}$}}
\newc{\wt}{\widetilde}

\def \lspone{\wt\chi_1^0}
\def \mlspone{m_{\lspone}}
\def \lsptwo{\wt\chi_2^0}

\newc{\ifb}{\mbox{${\rm fb}^{-1}$}}

\def \chonepm{\wt\chi_1^\pm}
\def \chonemp{\wt\chi_1^\mp}

\def \mchonepm{m_{\chonepm}}

\newc{\del}{\delta}

\def \lstop{\wt {t}_1}

\def \lsp{\wt\nu_{1,2}}
\def \mlsp{m_{\wt\nu_{1,2}}}
\begin{document}
\preprint{HIP-2018-15/TH, IP/BBSR/2018-10}
\title{Same-sign trilepton signal for stop quark in the presence of sneutrino dark matter}
\author{Dilip Kumar Ghosh}
\email{tpdkg@iacs.res.in}
\affiliation{Department of Theoretical Physics, Indian Association for the Cultivation of Science Jadavpur, Kolkata 700 032, India}
\author{Katri Huitu}
\email{katri.huitu@helsinki.fi}
\affiliation{Department of Physics, and Helsinki Institute of Physics, P. O. Box 64, FI-00014 University of Helsinki, Finland}
\author{Subhadeep Mondal}
\email{subhadeep.mondal@helsinki.fi}
\affiliation{Department of Physics, and Helsinki Institute of Physics, P. O. Box 64, FI-00014 University of Helsinki, Finland}
\author{Manimala Mitra}
\email{manimala@iopb.res.in}
\affiliation{Institute of Physics, Sachivalaya Marg, Bhubaneswar, Odisha 751005, India, Homi Bhabha National Institute, Training School Complex, Anushakti Nagar, Mumbai 400085, India}
\begin{abstract}
We have explored a minimal supersymmetric standard model scenario extended by one pair of gauge 
singlets per generation. In the model light neutrino masses and their mixings are generated via inverse seesaw 
mechanism. In such a scenario, a right-handed sneutrino can be the lightest supersymmetric particle 
and a cold Dark Matter (DM) candidate. 
If Casas-Ibarra parametrisation is imposed on the Dirac neutrino Yukawa coupling matrix ($Y_{\nu}$) to fit the 
neutrino oscillation data, the resulting $Y_{\nu}$ is highly constrained from the lepton flavor violating 
(LFV) decay constraints. 
The smallness of $Y_{\nu}$ requires the sneutrino DM to co-annihilate 
with other sparticle(s) in order to satisfy DM relic density constraint. We have studied 
sneutrino co-annihilation with wino and observed that this sneutrino-wino compressed parameter 
space gives rise to a novel same-sign trilepton signal for the stop quark, which is more 
effective than the conventional stop search channels in the present framework.   
We have shown that the choice of neutrino mass hierarchy strongly affects the signal event rate, 
making it easier to probe the scenario with inverted mass hierarchy. 
\end{abstract}
\maketitle
\section{Introduction}
\label{sec:intro}
Existence of non-zero neutrino masses and mixings \cite{Gonzalez-Garcia:2015qrr,deSalas:2017kay} remain one of the unsolved puzzles 
and one of the strongest motivations 
for looking for physics beyond the Standard Model (BSM). Supersymmetry (SUSY) remains one of the frontrunners among various BSM candidates. 
However, the R-parity conserving minimal supersymmetric standard model (MSSM) is unable to address the neutrino mass problem.
The easiest way to overcome this shortcoming is to incorporate a seesaw mechanism in the framework, where the light neutrinos gain 
non-zero masses through small mixing with additional gauge singlet fields. The canonical seesaw mechanism \cite{Minkowski:1977sc,
Yanagida:1979as,Mohapatra:1979ia,Glashow:1979nm,GellMann:1980vs} introduces three 
generations of gauge singlet superfields to the MSSM particle contents. The light neutrino mass scale leads to very small 
Dirac neutrino Yukawa couplings ($Y_{\nu}$) in the presence of TeV scale right-handed neutrinos. The smallness of $Y_{\nu}$ 
makes it difficult to get any observable effect of the neutrino sector at the LHC. Lepton number and (or) lepton flavor 
violating (LFV) decays on the other hand, can constrain these parameters further with increasing sensitivities. A more phenomenologically 
interesting option is provided by the inverse seesaw mechanism \cite{Mohapatra:1986aw,Mohapatra:1986bd,GonzalezGarcia:1988rw} 
where the $Y_{\nu}$ can in principle be as large as ${\mathcal O} (0.1)$ 
owing to the presence of a small lepton number violating parameter in the theory. Such large Yukawa parameters and sub-TeV RH neutrino mass 
can be constrained from collider as well as low energy experiments \cite{Das:2012ze,Das:2015toa,Das:2017nvm}. 

Presence of a right-handed (RH) neutrino superfield in a SUSY theory provides us with an exciting possibility of obtaining a RH 
sneutrino as the lightest SUSY particle (LSP) which can also be a good cold Dark Matter (DM) candidate \cite{Lee:2007mt,Cerdeno:2008ep,
Mondal:2012jv,BhupalDev:2012ru,DeRomeri:2012qd,Banerjee:2013fga,Ghosh:2014pwa,Chang:2017qgi,Chang:2018agk,DelleRose:2017ukx,DelleRose:2017uas,
DelleRose:2018mjj}. The left-handed 
sneutrino LSP option is strongly disfavored from DM direct detection constraint due to its gauge coupling with the $Z$-boson \cite{Falk:1994es,Hebbeker:1999pi}. 
A sub-TeV sneutrino LSP, in order to be considered as a DM candidate, therefore, has to be RH. This RH sneutrino arises from 
a singlet superfield and thus only couples to other particles via the Yukawa couplings.  The DM experimental data can be another probe 
to test the neutrino sector parameters in models augmented with gauge singlet superfield. 
The only pair annihilation process of any significance involves Higgs 
bosons in the s-channel, and unless the sneutrino mass is close to the scalar resonance region, the annihilation is not enough 
to produce correct relic density \cite{BhupalDev:2012ru,DeRomeri:2012qd,Banerjee:2013fga}. Thus a sneutrino DM mass around $\frac{m_h}{2}$, 
where $m_h$ indicates the 125 GeV Higgs boson mass, has been the favored region from the DM constraints. However, this region is 
now under severe scrutiny with the improvements in DM direct detection constraints. 

In the absence of the Higgs resonance region, it is quite difficult to produce enough annihilation cross-section for the RH sneutrinos 
in such scenarios. This forces us to look at co-annihilation options. With proper co-annihilation, a sneutrino can be a viable DM candidate  
throughout the mass range [100 GeV - 1 TeV]. However, direct search constraints on the other sparticles indirectly put constraint on the sneutrino 
masses in such cases. Attractive options involve winos or higgsinos as next to lightest SUSY particle (NLSP) since they always present 
the possibility of  co-annihilation with more than one particle simultaneously. Thus the DM constraints naturally leads to a compressed 
electroweak sector involving the LSP sneutrino and multiple neutralino-chargino states as the NLSP. As a result of this compression, the 
decay products of the NLSP neutralino-chargino are expected to be soft, resulting in weaker exclusion limits. This also means that such
a scenario will be hard to detect from direct production of the NLSP pairs. One can, however, produce them in a cascade and look for 
a multilepton channel with small Standard Model (SM) background to distinguish the signal.  

One such possibility can arise from stop quark production and its subsequent decay into the LSP sneutrino via the neutralino-charginos.
The physics of stop quark is of utmost importance in the MSSM-like theories that require to add a substantial correction to the tree level 
Higgs boson mass in order to increase it up to 125 GeV. This correction appears mostly from stop quark loop making the stop masses and 
mixing parameters of interest in the search of SUSY at the LHC. The existing constraints on the stop quark masses can extend up to 
1 TeV depending on its various decay modes \cite{Aaboud:2017dmy,Aaboud:2017ejf,Aaboud:2017nfd,Aaboud:2017wqg,Aaboud:2017ayj,
Aaboud:2017aeu,Aaboud:2018kya}. At the same time this exclusion limit can be relaxed in presence of a compressed 
electroweak sector as we have here. It turns out that the existing search channels are not sensitive enough to probe stop masses effectively 
under such circumstances owing to  the poor signal rates. Hence we have constructed a same-sign trilepton signal region, which is nearly 
background free and thus, despite of a poor event rate, can be used to probe TeV order stop masses at relatively low luminosity.  
We have also shown that the choice of normal hierarchy (NH) or inverted hierarchy (IH) in the light neutrino masses is reflected in the final 
event rate which draws a nice correlation between the neutrino and stop sector and also highlights a unique characteristic of such neutrino 
mass models.

The paper is organised as following. In section \ref{sec:model}, we briefly give an overview of the model and describe how neutrino 
oscillation data has been fit. In section \ref{sec:const} we discuss the phenomenological constraints on the model arising from LFV decays, 
DM and collider experiments. In section \ref{sec:coll} we discuss the canonical stop search strategies at the LHC and propose for the same a novel 
same-sign trilepton signal region, which is more suitable to probe the present scenario. Then we go on to define few benchmark 
points representative of the parameter space of our interest and perform a detailed collider analysis to present our results in the context 
of 13 TeV LHC. We have shown the exclusion limits on the stop mass derived from this study at moderate and high luminosities at the LHC 
in case of null result. In this context we have also presented the limits that can be expected at a high-energy (27 TeV) hadron machine. 
In section \ref{sec:concl} we summarise our results and conclude.
\section{Model }
\label{sec:model}
The supersymmetric inverse seesaw model (SISM) contains the SM gauge singlet superfields $\hat{N}$ and $\hat{S}$ with lepton numbers -1 and 
+1 respectively. The superpotential with these extra superfieds is
\begin{equation}
W=W_{\rm{MSSM}}+Y_{\nu} \hat{L}.H_u \hat{N}+M \hat{N} \hat{S}+ \mu_S \hat{S}.\hat{S}
\label{supis}
\end{equation}
In the above we consider three generations of $\hat{N}_i$ and $\hat{S}_i$ (i=1,2,3) and $\mu_S$ violates lepton number by two units 
($\Delta L=2$). 
The soft supersymmetry breaking Lagrangian for this model is
\begin{equation}
\mathcal{L}_{{soft}}=\mathcal{L}_{{MSSM}}-[m^2_N \tilde{N}\tilde{N}+m^2_S \tilde{S}\tilde{S}]-[A_{\nu}\tilde{L}{H_u}\tilde{N}+B_1 \tilde{N}\tilde{S}+B_2 \tilde{S}\tilde{S}+h.c]
\end{equation}

The $9 \times 9$ neutrino mass matrix in the basis $(\nu, N, S)$ has the following form, 
\begin{equation}
{\mathcal M}_{\nu}=\begin{pmatrix} 0 & M_D & 0 \\ M^T_D & 0 & M \\ 0 & M^T & \mu_S \end{pmatrix}
\label{matrix}
\end{equation}
where $M_D$, $M$ and $\mu$ are $3\times 3 $ matrix and $M_D= Y_{\nu} v_u$, where $v_u$ is the vacuum expectation value (VEV) 
of $H_u$. For the parameter $||\mu_S|| << ||M||$, the light neutrino mass matrix becomes, 
\begin{equation}
M_{\nu} \sim M_D{M^T}^{-1} \mu_S M^{-1}M^T_D
\label{mnu}
\end{equation}
The matrix $\mu_S$ can be small\footnote{Note that, $\mu_S$ breaks lepton number by two units. Hence in the limit $\mu_S\to 0$, the symmetry of the theory
enhances. From the naturalness argument, $\mu_S$ is preferred to be small. Moreover, a small $\mu_S$ is absolutely
essential to incorporate inverse seesaw mechanism. Moreover, a large $\mu_S$ will cause the $\Delta L =2$ processes 
like neutrinoless double beta decay rate to increase rapidly \cite{Deppisch:2015qwa}.} enough to explain the eV neutrino mass constraint. 
In passing we would like to add a few comments on the choice of our superpotential. Since we want the lightest right-handed  
sneutrino state to be a stable LSP and hence a DM candidate, we choose to work within R-parity conserving framework. 
This choice prevents us to write terms violating lepton number by odd unit(s),e.g., $\hat N\hat N\hat N$, $\hat N\hat N\hat S$, 
$\hat S\hat S\hat S$ and $\hat N\hat S\hat S$. We do not allow 
the scalar components of $\hat N$ and $\hat S$ to obtain VEVs for the same reason. However, one can in principle, 
allow additional $\Delta L=2$ terms, i.e., non-zero $\hat L\hat S$ and $\hat N\hat N$ terms in the superpotential. 
However, the presence of the $\hat N\hat N$ term does 
not affect the inverse seesaw structure at tree-level as the rank of $\mathcal{M}_{\nu}$ remains same. On the other hand, 
the $\hat L\hat S$ term 
will modify eq.~\ref{mnu} unless the coupling $Y_S$ in $Y_S\hat L.\hat H_u\hat S$ is very small, $Y_S\lesssim 10^{-12}$. One can impose additional 
symmetry \cite{Khalil:2010iu,Dev:2009aw} in the superpotential to forbid such terms. We have chosen to work within a minimal 
set up \cite{Deppisch:2004fa,Hirsch:2009ra,Abada:2012cq,Abada:2014kba} that can address the neutrino oscillation data via inverse seesaw mechanism.  

The light neutrino mass matrix is diagonalized by the PMNS mixing matrix $U_{\nu}$ as follows. 

\begin{equation}
U_{\nu}^T M_{\nu} U_{\nu}^*=M^d_{\nu}
\label{lightmass}
\end{equation}
We consider the matrices $M$ and $\mu_S$ to be diagonal. One can follow the Casas-Ibarra parametrization \cite{Casas:2001sr} to construct  the matrix $R$
\begin{equation}
R={M^d}^{-1/2}_{\nu}U^T_{\nu}M_D{M^T}^{-1}{\mu_S}^{1/2}, 
\label{casas}
\end{equation}
where  $R$ is a complex orthogonal matrix with $RR^T=I$. Throughout the paper, we consider the best fit values of the oscillation 
parameters \cite{deSalas:2017kay}. The Dirac mass matrix $M_D$ can be fixed in terms of the light neutrino masses, PMNS mixing matrix and $R$ as
\begin{equation}
M_D=U^*_{\nu}\sqrt{M^d}_{\nu}R \mu_S^{-1/2}M
\label{eq:R}
\end{equation}
In our subsequent discussion, we consider the simplest scenario with $R=\mathbb{I}$.  The heavy  neutrino masses  are proportional to $ M \pm \mu_S$. 
After diagonalization, this gives rise to the quasi-degenerate sterile neutrino states with mass splitting $\delta M \sim \mu_S$. 

Below, we discuss a number of experimental constraints arising from the lepton flavor violating searches, dark matter searches and collider constraints. 
\section{Experimental Constraints}
\label{sec:const}
In this section, we discuss different experimental constraints on the heavy neutrino and sneutrino parameters in addition to the neutrino 
oscillation data. The heavy neutrinos in the inverse seesaw can give large contribution in  the loop mediated LFV process 
$\ell_i \to \ell_j \gamma$, $\ell_i\to\ell_j\ell_j\ell_j$ and $\ell_i \to \ell_j$ conversion in nuclei \cite{Deppisch:2004fa,Hirsch:2009ra,
Abada:2012cq,Sun:2013kga,Abada:2014kba,Marcano:2017ucg}. 
Additionally, for 
much lower masses upto few GeV, the sterile neutrinos can further be constrained from different LFV meson decays $M \to M^{\prime} l_i l_j$, 
fixed target experiments, the peak searches in $\pi \to e \nu$ and $K \to e \nu$, as well as $\beta$ decay \cite{Lello:2012gi}. 
Active-sterile neutrino mixing upto $\theta^2 \sim 10^{-11}$ can be probed in future FCC-ee \cite{Blondel:2014bra} and  SHiP \cite{Anelli:2015pba}, 
relevant for  few GeV mass range of the RH neutrino.  
SISM is augmented with small lepton number violation owing to the smallness of $\mu_S$. Therefore, the LNV processes,  
such as $0\nu 2 \beta$, LNV meson decays $K \to \pi l^{\pm} l^{\pm}$ will be suppressed in  this model. We consider the heavy neutrinos 
in the [100 GeV - 1 TeV] range that can be most optimally tested through the flavor violation and collider searches. In the SISM, 
$\ell_i \to \ell_j \gamma$ and $\ell_i \to \ell_j$ conversion in 
nuclei  are further enhanced due to  sneutrino-chargino and slepton-neutralino mediated loop contributions.  
We furthermore assume that the sneutrino is the LSP and obtain the constraints from the dark matter relic density. 

\subsubsection*{LFV Constraints}
\label{sec:const_lfv}
As an artefact of fitting the neutrino oscillation data as described in the previous section, the resulting Dirac neutrino mass matrix, $M_D$, 
becomes off-diagonal which can potentially enhance LFV decays like $\ell_i\to\ell_j\gamma$. 
We keep the Yukawa matrix for the 
charged lepton diagonal. As a result, the LFV decays are generated at one loop through contributions from lepton-neutrino-W boson, lepton-slepton-neutralino and 
lepton-sneutrino-chargino loops. Both these non-supersmmetric and supersymmetric loop contributions have been studied in detail \cite{Deppisch:2004fa,Abada:2011hm, 
Abada:2014kba}. 
Depending on the choices of the neutrino sector parameters, namely, 
$M_D$, $M$, $\mu_S$ and the chargino and sneutrino masses, these LFV decay rates can be significant. The experimental limits on such LFV decay 
branching ratios \cite{Adam:2013mnn,Baldini:2013ke,Aubert:2009ag,Aushev:2010bq,Hayasaka:2010np,Bertl:2006up,Bartoszek:2014mya} thus can 
restrict the model parameter space effectively. The existing constraints are more stringent for LFV decays of muons than  
of taus owing to better detection efficiency of the former. Hence stringent restrictions on the $Y_{\nu}$ parameters are to be expected when 
inverted hierarchy among the light neutrino masses are considered owing to the resultant large $Y_{\nu}$ corresponding to the first generation. 
In order to study the impact of these constraints on the LSP mass and its 
couplings, we have performed a scan where $M$ has been varied within the range [100 GeV - 1 TeV]. Wino mass parameter $M_2$ has been adjusted 
in such a way that $\mlsp < \mchonepm \le \mlsp + 50$ GeV to facilitate co-annihilation with the LSP, which has been discussed in the subsequent 
subsection. The value of tan$\beta$ has been kept fixed at 10 and different $\mu_S$ values ($10^{-8}$, $10^{-7}$ and $10^{-6}$ GeV) have been 
chosen with $Y_{\nu}$ fitted according to Eq.~\ref{eq:R}. The computation of particle masses, mixing and their decay have been performed 
using SPheno\footnote{All the masses are calculated with corrections upto one loop, apart from the Higgs mass which is computed upto two loop level.} 
\cite{Porod:2003um,Porod:2011nf,Porod:2002wz} after model implementation using SARAH \cite{Staub:2008uz,Staub:2009bi,Staub:2010jh,
Staub:2013tta,Staub:2015kfa}. 
For the LFV decay branching ratio calculations we have used the results from ref.~\cite{Abada:2014kba} which include both the SUSY 
and non-SUSY contributions.

\begin{figure}[h!]
\centering
\includegraphics[scale=0.2]{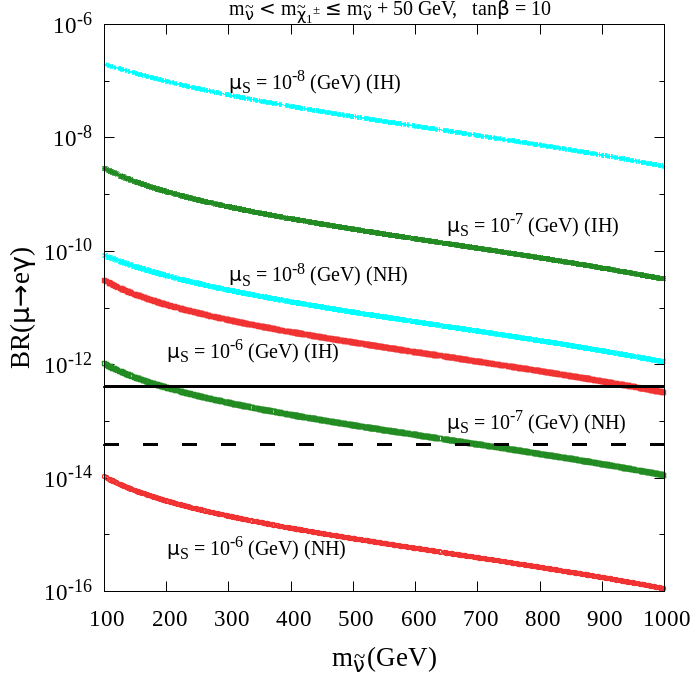}
\includegraphics[scale=0.2]{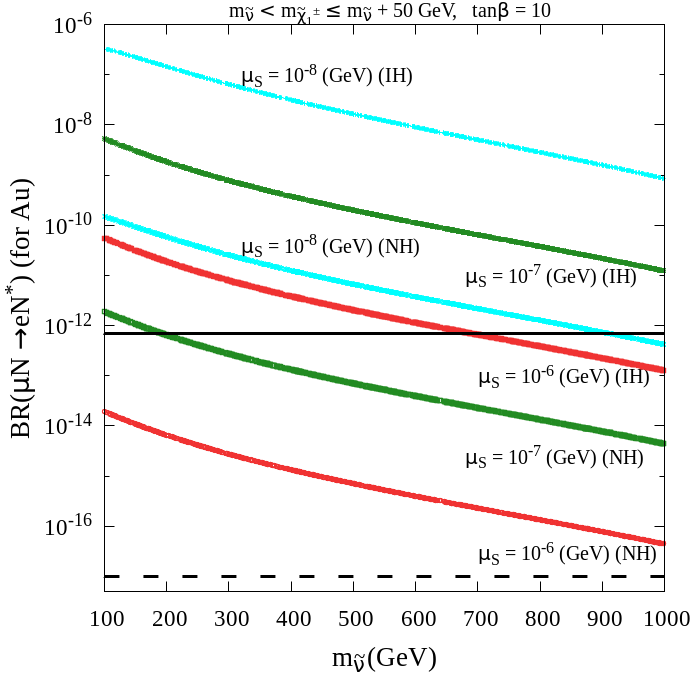} 
\includegraphics[scale=0.2]{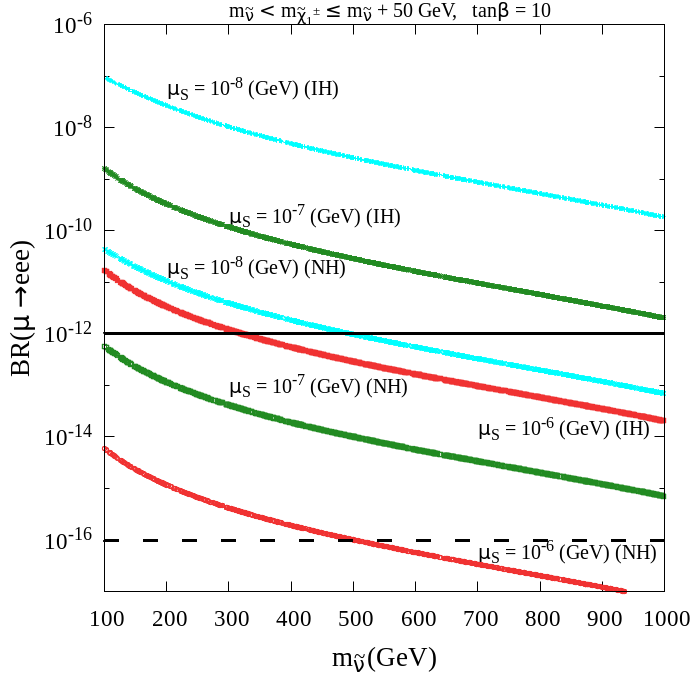} \\
\includegraphics[scale=0.2]{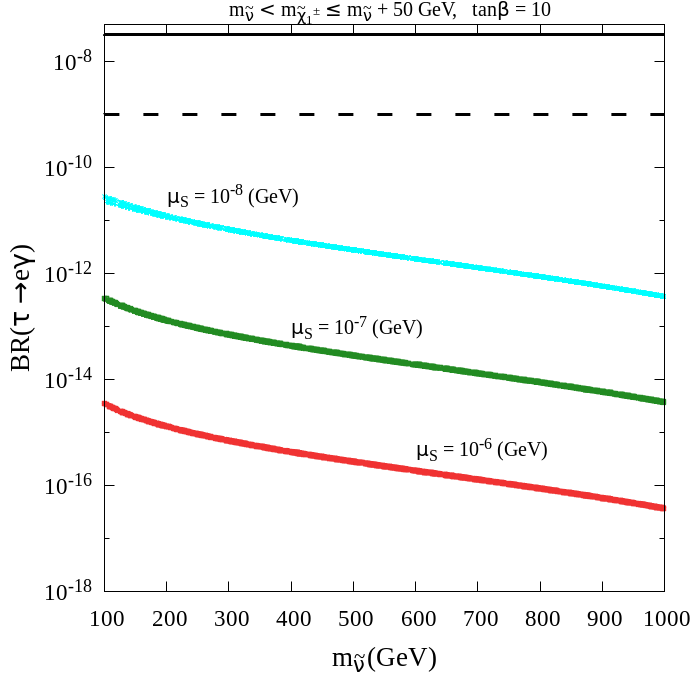}
\includegraphics[scale=0.2]{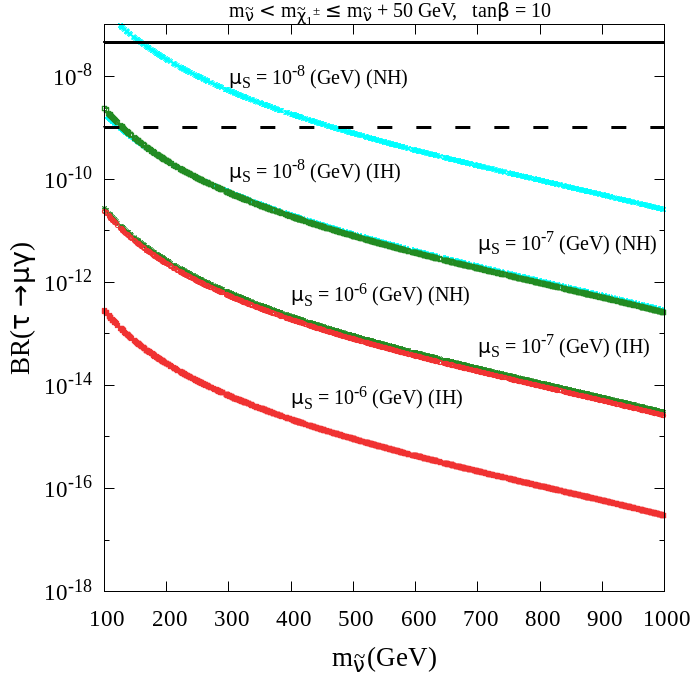}
\includegraphics[scale=0.2]{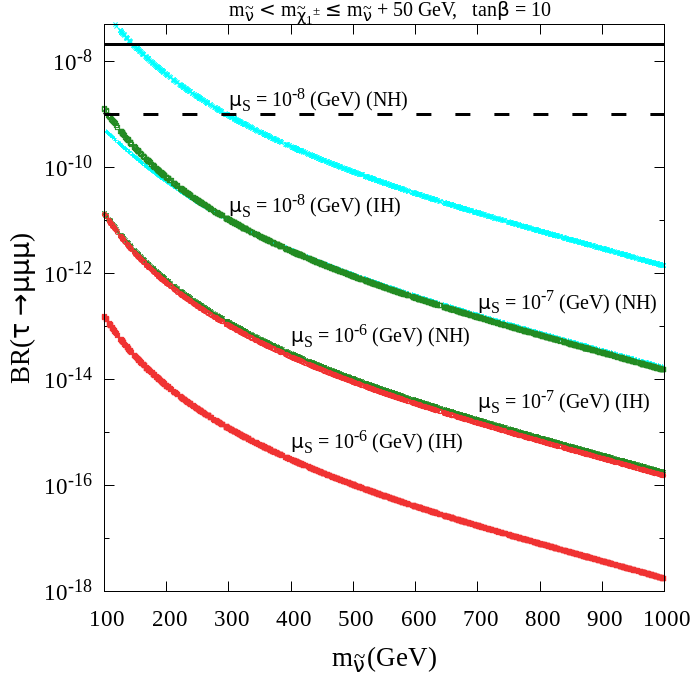}
\caption{Top panel:  the predicted branching ratio of $\mu \to e \gamma$, $\mu-e$ conversion in nuclei and $\mu \to eee$ vs the sneutrino mass. 
Lower panel: the branching ratio for $\tau \to e \gamma$, $\tau \to \mu \gamma$ and $\tau \to \mu \mu \mu$. 
The horizontal solid black line corresponds to 
the present experimental bounds on BR$( \mu \to e\gamma) \leq 4.2 \times 10^{-13}$, BR$(\mu N \to eN^*) \leq 10^{-12}$, 
BR$( \mu \to eee) \leq 10^{-12}$, BR$(\tau \to e\gamma) \leq  3.3 \times 10^{-8}$,
BR$(\tau \to \mu\gamma) \leq  4.4 \times 10^{-8}$ and BR$(\tau \to \mu \mu \mu) \leq 2.1 \times 10^{-8}$ 
\cite{Adam:2013mnn,Baldini:2013ke,Aubert:2009ag,Aushev:2010bq,Hayasaka:2010np,Bertl:2006up,Bartoszek:2014mya}. 
The corresponding future sensitivities are shown by the horizontal dotted black lines \cite{Adam:2013mnn,Baldini:2013ke,Aubert:2009ag,Aushev:2010bq,Hayasaka:2010np,Bertl:2006up,Bartoszek:2014mya}. }
\label{fig:lfv}
\end{figure}

Fig.~\ref{fig:lfv} shows the variation of the relevant LFV decay branching ratios as a function of the LSP mass $\mlsp$ \footnote{Heavy neutrino masses 
are same as the sneutrino masses since they are both driven by the same parameter $M$.}. The different colored lines correspond to different choices of 
$\mu_S$ while the black solid and dotted lines represent the present and future experimental sensitivities for the respective LFV processes. As expected, the 
constraints are most severe for the IH case in the $\mu\to e\gamma$ and $\mu N\to eN^*$ modes. 
%
In particular, for IH in the light neutrino masses, the LNV parameter  $\mu_S < 10^{-6}$ GeV is excluded 
throughout the entire LSP mass region upto $\tilde{m}_{\nu} =1000$ GeV. 
The restriction is relatively much weaker from LFV $\tau$ decays, $\tau \to e \gamma$, $\tau \to \mu \gamma$ and $\tau \to \mu \mu \mu$. 
The experimental limits on these 
decay modes are  yet to improve significantly to probe the parameter space in consideration. 
As expected, for the $\tau$ LFV decays, NH mass hierarchy predicts a larger branching ratio 
than the IH mass hierarchy because of the largeness in third generation Yukawa couplings. This is easily testable in the next generation experiments.  
With the recent results of the neutrino oscillation experiment No$\nu$a  disfavoring IH in neutrino sector \cite{Adamson:2017gxd,NOvA:2018gge},  
the predictions for NH are even more significant. For $\tau \to e\gamma$ the predicted branching 
ratio is smaller by more than $\mathcal{O}(10^{4})$  than $\tau \to \mu\gamma$.
In this case, both the lines corresponding to NH and IH merge to result in one unique line for each $\mu_S$ values 
\footnote{In the distributions of $BR(\tau\to\mu\gamma)$ and
$BR(\tau\to\mu\mu\mu)$, the IH lines corresponding to the smaller $\mu_S$ overlaps with the NH line corresponding to the subsequent $\mu_S$. 
The small numerical difference between the lines is indistinguishable in the logarithmic scales of the figures.}.
The restriction from $\tau \to eee$  is the weakest, and hence we do not explicitly show that.  
As the future sensitivities show, $\mu N\to eN^*$ conversion process is most likely to probe this parameter space entirely.

Apart from the neutrino sector parameters, the masses of the neutralino and charginos can also be constrained from these LFV 
decay computations. In order to get Fig.~\ref{fig:lfv}, we have only kept the wino mass close to the LSP as mentioned above. All other neutralino-chargino 
masses are kept above 2.0 TeV. Since the wino pair lie close to the LSP mass, some of the parameter region with light enough sneutrino mass can be ruled 
out from the LHC data. However, LFV constraints on these masses depend on the choice of $\mu_S$. As shown in Fig.~\ref{fig:lfv}, large $\mu_S$ results in 
weaker LFV constraint. We checked that for $\mu_S\sim 10^{-5}$ GeV, even for more constraining inverted hierarchy scenario, sneutrino and chargino masses 
close to 200 GeV are still allowed from the LFV decays. However, that parameter space will be ruled out from LHC direct search constraints on the gaugino masses.
Note that, LFV decay rates are not significant enough to constrain SUSY particle masses above  100 GeV for $\mu_S\gtrsim 10^{-6}$ GeV if the light neutrino masses 
are aligned in normal hierarchy.
\subsubsection*{Dark Matter Constraints}
\label{sec:const_dm}
In the SISM, a light RH sneutrino cold DM candidate can be ideally fit in the Higgs resonance region 
\cite{BhupalDev:2012ru,DeRomeri:2012qd,Banerjee:2013fga}, i.e., $\mlsp$ requires to lie in the vicinity of $m_h/2$, where $m_h$ is the SM-like
Higgs mass in order to ensure enough annihilation to satisfy the relic density constraint \cite{Hinshaw:2012aka}. 
The RH sneutrino states are required to have sufficiently small mixing with the LH sneutrino states in order to avoid the direct detection 
cross-section ($\sigma_{SI}$) constraints. However, the most recent constraint on $\sigma_{SI}$ imposed by DM experiments like XENON1T 
and PANDA \cite{Aprile:2017iyp,Cui:2017nnn,Aprile:2018dbl} have already excluded this parameter space. 
On the other hand, imposing the LFV constraints on the LSP mass and 
couplings renders the $Y_{\nu}$ parameters sufficiently small so that the resulting $\sigma_{SI}$ 
in the sub-TeV $\lsp$ mass region lies orders of magnitude below the present experimental sensitivity 
\cite{Aprile:2017iyp,Cui:2017nnn,Aprile:2018dbl}. However, such small $Y_{\nu}$ implies that 
even the Higgs resonance region can not provide sufficient annihilation for the 
RH-sneutrino LSP to satisfy the relic density constraint. As a consequence, one has to look for co-annihilation of $\lsp$ with other sparticles. 
Here we have studied the sneutrino co-annihilation  with wino-like chargino and neutralino. The right-handed LSP sneutrino can not interact 
directly with the wino components of the chargino-neutralino pair and can only do so through their higgsino component. However, we have studied 
the case where the NLSP chargino-neutralino pair are almost purely wino-like. Under this circumstance, the resultant relic density is entirely due to the 
co-annihilation of the chargino-neutralino pair. The contribution of sneutrino annihilation to the relic density is negligibly small \cite{Arina:2013zca}. Hence, the LSP 
can be purely right-handed, that results in elusively small direct detection cross-section.  
We have use MicrOmegas \cite{Belanger:2013oya} for the calculation of relic density and direct detection cross-section.

It is worth mentioning that a higgsino NLSP can also be a good choice. Due to the presence of three nearly degenerate neutralino-chargino 
states that can co-annihilate with the sneutrino, a higgsino NLSP scenario is also capable of producing the correct relic density with 
appropriate choice of the higgsino parameters. A bino LSP scenario, on the other hand, fail to meet the relic density criteria 
due to lack of sufficient co-annihilation channels. 
\begin{figure}[h!]
\centering
\hspace{-1cm}
\includegraphics[scale=0.3]{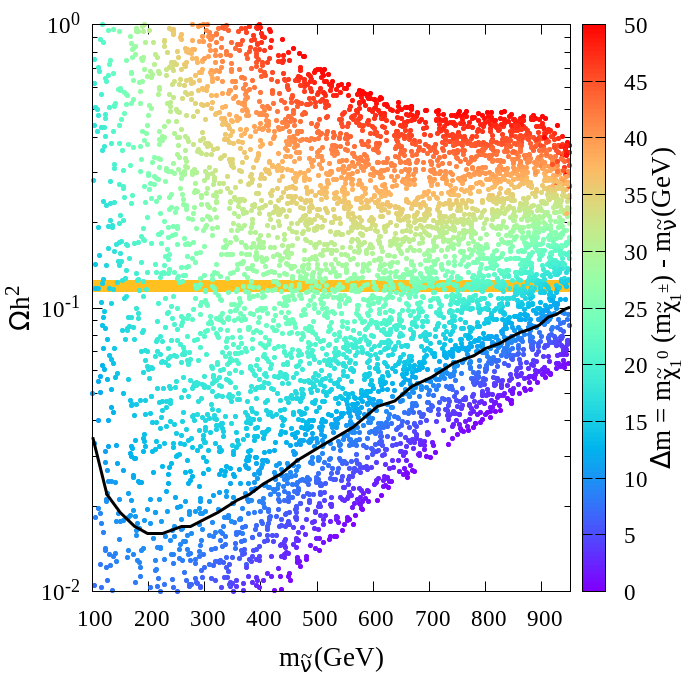}
\caption{The variation of the relic density is shown as a function of the LSP sneutrino mass. 
The color coding represents the LSP-NLSP mass gap, $\Delta m$. 
The horizontal band represents  $2\sigma$ allowed range of relic density, $\Omega h^2 = 0.119\pm 0.0027$ \cite{Hinshaw:2012aka}.
The black contour represents $\Delta m\sim 10$ GeV.}   
\label{fig:relic}
\end{figure}

Fig.~\ref{fig:relic} shows the relic density ($\Omega h^2$) distribution as a function of DM mass, $\mlsp$ after taking into account 
neutrino mass and LFV constraints. tan$\beta$ is kept fixed at
a moderate value, 10\footnote{Choice of tan$\beta$ does not impact the sneutrino mass range or the LSP-NLSP mass gap required to produce 
correct relic density.}
and the $Y_{\nu}$ matrix is derived at each point according to Eq.~\ref{eq:R}. The NH or IH choices do not have significant impact on 
the distribution. However, we keep $\mu_S=10^{-5}$ GeV in order to be safe from the LFV constraints. 
The parameter $M$ has been varied in the range [100 - 1000] GeV as before and to ensure co-annihilation the LSP-NLSP mass difference 
($\Delta m = \mlspone (\mchonepm) - \mlsp$) have been kept within 50 GeV by adjusting the wino mass 
parameter, $M_2$. The $\Delta m$ 
variation is color coded according to the gradient bar on the right. The horizontal golden shaded region indicates the $2\sigma$ allowed 
range of relic density, $\Omega h^2 = 0.119\pm 0.0027$ \cite{Hinshaw:2012aka}. As evident from the figure, $15 < \Delta m < 30$ GeV 
is favored from $\Omega h^2$ consideration. Therefore, one can derive stringent constraint on the NLSP masses 
subjected to the sneutrino DM mass from relic density requirement. The black contour in Fig.~\ref{fig:relic} shows the boundary below which 
$\Delta m < 10$ GeV. With such small $\Delta m$, the decay products of the NLSP will practically be undetectable. However, in this work, 
we are only concerned with the points lying over the relic density allowed band. Given the allowed $\Delta m$ corresponding to these 
points, the resulting NLSP decay products will be soft but detectable with varied efficiency \footnote{In our collider analysis, in order 
to attain maximal cut efficiency, we have considered $\Delta m\sim 25$ GeV.}.
\subsubsection*{Collider Constraints}
\label{sec:const_coll}
Since the sneutrino DM requires co-annihilation with a wino in the present scenario, the existing 
limits on the mass of wino-like chargino and neutralino can further constrain the DM allowed parameter region indirectly. The most stringent 
constraints from LHC on the wino mass is derived from trilepton and(or) dilepton final states resulting from $\lsptwo\chonepm$ and $\chonepm\chonemp$ 
production channels, respectively, with the assumption that $\lsptwo$ and $\chonepm$ are wino-like and mass degenerate while the $\lspone$ 
is the LSP and is bino-like \cite{Aaboud:2018jiw,Sirunyan:2018ubx,Aaboud:2017leg,Aaboud:2017mpt,Sirunyan:2017lae,Aaboud:2017nhr,Sirunyan:2017zss}. 
However, in our scenario the wino-bino mass hierarchy is reversed and sneutrino being the LSP, the final decay products are 
kinematically quite different.
The co-annihilation requirement puts the wino $\lspone$-$\chonepm$ pair close to the $\lsp$ mass and as a result, the leptons 
arising from the $\chonepm$ decay are quite soft. $\lspone$ on the other hand, decays completely invisibly. Besides, the $\lsptwo\chonepm$ 
production cross-section is much smaller here compared to that used by the experimental analyses because of the sizable mass gap between the two states. 
$\chonepm\lspone$ production cross-section is expected to be larger, but the mono-lepton signal has a huge background contribution arising from 
single W production channel at the LHC. Thus the neutralino-chargino production channels result in much weaker limits on the relevant sparticle masses. 
Existing constraints from $\chonepm\chonemp$ production can also restrict the SISM scenario. However, the resultant dilepton signal region 
requires relatively large lepton $p_T$ cuts that reduces the signal event rate drastically in our case. One can revisit the dilepton 
analyses of ATLAS and CMS in this regard with softer $p_T$ cuts on the leptons. Nevertheless, one has to use a large stransverse mass ($M_{T2}$) cut  
in order to get rid of the background arising from $W$ boson pair production. This cut discards most of the signal events from chargino pair production
here because of the small chargino-sneutrino mass gap. Moreover, the absence of large transverse missing energy makes this channel
less sensitive at the LHC. The existing exclusion limits are clearly not sensitive when LSP-NLSP mass gap is $\sim 20$ - 30 GeV 
even with small $M_{T2}$ cuts \cite{Aaboud:2018jiw}.
All these factors combine to make the existing limits on the gaugino masses much more relaxed in our case and allow us to choose sufficiently light 
sneutrino DM masses. We have ensured that the existing constraints on the gaugino masses are properly taken into account by testing our 
parameter space against the LHC results via CheckMATE \cite{Drees:2013wra,Dercks:2016npn}.  
\section{Collider Phenomenology}
\label{sec:coll}
Our discussion of the SISM scenario and the impact of the neutrino oscillation, LFV and DM experimental constraints point towards a naturally compressed 
spectrum consisting of the LSP sneutrino and the NLSP wino-like neutralino-chargino pair. As discussed in the previous section, 
the canonical search strategy for the gauginos at the LHC may not be sensitive enough to probe this scenario or distinguish the 
signal from that expected from a MSSM compressed spectra. We observe that a clean and distinguishable signal region for our 
scenario can be obtained via secondary production of the gauginos from a stop cascade. Note that, although the choices of wino 
masses in this framework are subjected to the choice of the LSP mass, the bino-like neutralino ($\lsptwo$) and the colored 
sparticle masses lie anywhere as long as they are not excluded by the LHC data.

The stop quark search at the LHC \cite{Aaboud:2017ejf,Aaboud:2017nfd,Aaboud:2017wqg,Aaboud:2017ayj,Aaboud:2017aeu,Aaboud:2018kya} 
mostly concentrates on the two stop decay modes $\tilde t\to t\lspone$ and $\tilde t\to\chonepm b$, where $\lspone$ and $\chonepm$ 
are bino and wino-like, respectively. The various signal regions include 0,1 or 2 lepton final states associated with b-jets 
and missing transverse energy ($\met$). The most stringent constraint on the stop masses in the MSSM framework is derived assuming 
either $\tilde t\to t\lspone$, $\tilde t\to b\chonepm (\chonepm\to W\lspone)$ or $\tilde t\to bff^{\prime}\lspone$, where $f$ refers to the fermions. 
The existing bound on the stop masses can extend upto 1 TeV for a massless LSP \cite{Aaboud:2017ayj,Aaboud:2017aeu}. Although 
similar final states can be obtained in our scenario, these conventional decay branching ratios can be quite small and at the same 
time, the leptons originating from the NLSP decay $\chonepm\to\ell\lsp$ are 
expected to be soft and consequently result in worse cut efficiency. Thus the conventional search strategies for the lighter stop 
may not be applicable to probe this scenario. 

A very clean and unique final state can result from right-handed stop pair production and subsequent decay of a stop into a bino-like 
$\lsptwo$ along with a top quark, while $\lsptwo$ further decays to the wino-like chargino $\chonepm$ 
and a $W$-boson. The $\chonepm$ finally decays to a charged lepton and LSP sneutrino. Owing to the Majorana nature of the $\lsptwo$, 
this cascade can result at LHC in a same-sign tri-lepton signal, which is almost background free. 
The full-decay chain that we consider here is, therefore, as follows: \\
\begin{eqnarray}
p p \to \lstop  \lstop^* \to t \lsptwo \bar{t} \lsptwo \rightarrow b W^{+} \bar{b} W^{-} \chonepm \chonepm W^{\mp} W^{\mp}; 
~~\chonepm\rightarrow\ell^{\pm}\lsp \nonumber 
\end{eqnarray}
Final state consists of  $b \bar{b} + {\rm n-jets} + \ell^{\pm} \ell^{\pm} \ell^{\pm}+\cancel{E_T} ~~~(\ell\equiv e, \mu)$, 
where two of the same-sign leptons 
originate from the two charginos while the third originates from one of the $W$-bosons arising from top quark decays. Note that, 
a similar hierarchy in the stop and gaugino masses can be obtained in the MSSM as well with wino-like LSP. However, in that case, 
in order to result in a same-sign trilepton final state, three of the same-sign $W$-bosons in the cascade would have to decay 
leptonically. The final event rate is, therefore, expected to be less \cite{Aaboud:2017dmy} due to the small $W$ leptonic decay 
branching ratio. 
However, in our present scenario, the NLSP $\chonepm$ can only decay leptonically ($e$, $\mu$ or $\tau$) into the sneutrino and 
only one of the $W$s in the cascade is required to do the same. 
Hence the event rate is expected to be much larger. On the other hand, the LSP-NLSP compressed region ($\Delta m\simeq 25 - 30$ GeV) 
results in softer leptons, which 
affects the cut efficiency and dents the event rate somewhat, but the same-sign trilepton being a clean channel, proves to be 
much more effective in probing the SISM parameter space than the conventional signal regions. 

\subsection{Sample benchmark points }
\label{sec:bps}
In Table~\ref{tab:bp_br} below, we have presented the relevant parameters, masses, branching ratios and different experimental 
constraints corresponding to two sample benchmark points assuming NH and IH in each cases.
\begin{table}[h!]
\begin{center}
\scriptsize
\begin{tabular}{||c||c|c||c|c||}
\hline
\multicolumn{1}{||c||}{Parameters, masses \&} &
\multicolumn{2}{|c||}{\bf BP1} &
\multicolumn{2}{|c||}{\bf BP2}\\
\cline{2-5}
Branching Ratios & Normal & Inverted & Normal & Inverted \\
\hline\hline
$M_1$ (GeV) & 530.0& 530.0& 740.0& 740.0\\
$M_2$ (GeV) & 401.1& 401.1& 595.8& 595.8\\
$M^2_{\tilde t_L}$ (${\rm GeV}^2$) & $6.0\times 10^6$& $6.0\times 10^6$& $6.0\times 10^6$& $6.0\times 10^6$\\
$M^2_{\tilde t_R}$ (${\rm GeV}^2$) & $5.5\times 10^5$& $5.5\times 10^5$& $8.0\times 10^5$& $8.0\times 10^5$\\
$\mu$ (GeV) & 1500.0& 1500.0& 1500.0& 1500.0\\
$A_t$ (GeV) & -2200.0& -2200.0& -2200.0& -2200.0\\
tan$\beta$ (GeV) & 10.0& 10.0& 10.0& 10.0\\
$M^{11}$ (GeV) &400.0 &400.0 &600.0 &600.0 \\
$\mu^{ii}_S$ (GeV) &$10^{-5}$ &$10^{-5}$ &$10^{-5}$ &$10^{-5}$ \\
\hline\hline
$Y_{\nu} (\times 10^2)$ & {\tiny $\left(\begin{array}{ccc}
0.019 & -0.264 & 0.362 \\
0.013 & 0.303 & -0.805 \\
0.003 & 0.358 & 0.949\\ 
\end{array}\right)$} & {\tiny $\left(\begin{array}{ccc}
0.425 & -0.636 & 0.016 \\
0.281 & 0.730 & -0.036\\
0.076 & 0.862 & 0.042\\
\end{array}\right)$} & {\tiny $\left(\begin{array}{ccc}
0.029 & -0.264 & 0.362 \\
0.019 & 0.303 & -0.805 \\
0.005 & 0.357 & 0.949 \\ 
\end{array}\right)$} & {\tiny $\left(\begin{array}{ccc}
0.637 & -0.636 & 0.016 \\
0.421 & 0.730 & -0.036 \\
0.114 & 0.862 & 0.042 \\
\end{array}\right)$} \\
\hline\hline
$m_{\tilde t_1}$(GeV) & 860.1& 860.1& 994.8& 994.8\\
$m_{\tilde\chi^{\pm}_1}$(GeV) & 426.7& 426.7& 627.3& 627.3\\
$m_{\tilde\chi^0_2}$(GeV) & 530.9& 530.9& 741.6& 741.6\\
$m_{\tilde\chi^0_1}$(GeV) & 426.5& 426.5& 627.1& 627.1\\
$m_{\tilde\nu_{1,2}}$ (GeV) & 400.0& 400.0& 600.0 & 600.0 \\
$m_N$ (GeV) &  400.0&  400.0& 600.0 & 600.0\\
\hline
BR($\tilde t_1\to\tilde\chi^0_2 t$) & 0.89  & 0.89  & 0.84 & 0.84 \\
BR($\tilde t_1\to\tilde\chi^0_1 t$) & 0.03  & 0.03  & 0.05 & 0.05 \\
BR($\tilde t_1\to\tilde\chi^{\pm}_1 b$) & 0.08  & 0.08  & 0.11 & 0.11 \\
BR($\tilde\chi^0_2\to\tilde\chi^{\pm}_1 W^{\mp}$) & 0.99  & 0.99  & 0.99 & 0.99 \\
BR($\tilde\chi^{\pm}_1\to\tilde\nu_{1,2}e$) & -  & 0.31 & - & 0.50 \\
BR($\tilde\chi^{\pm}_1\to\tilde\nu_{1,2}\mu$) & 0.35 & 0.69 & 0.35 & 0.50 \\
BR($\tilde\chi^{\pm}_1\to\tilde\nu_{1,2}\tau$) & 0.65 & - & 0.65 & - \\
\hline
BR($\mu\to e\gamma$) & $4.5\times 10^{-18}$ & $1.3\times 10^{-14}$ & $1.4\times 10^{-18}$ & $4.0\times 10^{-15}$ \\
BR($\mu N\to eN $) (Au) & $5.1\times 10^{-18}$ & $1.5\times 10^{-14}$ & $1.1\times 10^{-18}$ & $3.3\times 10^{-15}$ \\
BR($\tau\to e\gamma$) & $1.5\times 10^{-18}$ & $1.5\times 10^{-18}$ & $4.6\times 10^{-19}$ & $4.5\times 10^{-19}$ \\
BR($\tau\to\mu\gamma$) & $2.9\times 10^{-16}$ & $3.4\times 10^{-18}$ & $3.9\times 10^{-17}$ & $4.5\times 10^{-19}$ \\
\hline
$\Omega h^2$(GeV) &0.120 & 0.120 & 0.120 & 0.120 \\
$\sigma_{SI}$ (pb) & $8.8\times 10^{-18}$ & $8.8\times 10^{-18}$ & $2.0\times 10^{-19}$ & $2.0\times 10^{-19}$ \\
\hline\hline
$pp\to\tilde t_1\tilde t_1$(fb) &17.55 &17.55 &6.38 &6.38 \\
\hline\hline
\end{tabular}
\caption{Relevant parameters, masses and branching ratios along with relic density, direct detection cross-section and stop 
pair production cross-section at 13 TeV LHC corresponding to the two benchmark points for both NH and
IH of light neutrino masses. $\mu^{ii}_S$ represents the three diagonal entries of the matrix $\mu_S$ and $M^{11}$ represents the 
first generation diagonal entry in the matrix $M$. The other two diagonal entries $M^{22}$ and $M^{33}$ are kept fixed at 1000 GeV.}
\label{tab:bp_br}
\end{center}
\end{table}

For our choice of the sneutrino mass and lepton number violating parameter $\mu_S$, the LFV branching ratios are well within 
the present experimental limits, and the IH case of BP1 is the one closest to be probed by BR$(\mu\to e\gamma)$ in near future. 
The DM direct detection cross-sections are also rendered quite small. The choice of NH or IH is highlighted in the resulting 
$Y_{\nu}$ matrix. The larger third generation $Y_{\nu}$ results in larger decay BR of the $\chonepm$ into $\tau\lsp$ in the 
NH case. On the other hand, in the IH case, larger first generation $Y_{\nu}$ leads to a larger BR($\chonepm\to e\lsp$). Naturally, 
one would expect larger leptonic event rates in the IH cases compared to the NH ones when the lighter chargino appears in the 
cascade. The $\tilde t_1$, being mostly right-handed, decays dominantly into a top quark and the bino-like $\lsptwo$. In the absence
of the $\lspone h$ mode, $\lsptwo$ decays entirely into $\chonepm W^{\mp}$. The mass difference between $\lsptwo$ and $\lspone$  
is chosen to ensure this so that the largest possible signal rate can be estimated. 

Note that, the key difference in between {\bf BP1} and {\bf BP2} lies in the choices of the lighter stop and the LSP sneutrino mass. 
The relevant branching ratios of the $\lsptwo$ and $\tilde t_1$ are similar in both the benchmark points. BR$(\chonepm\to\ell\lsp)$, 
where $\ell=e~{\rm or}~\mu$, slightly differ in the IH case due to the resultant $Y_{\nu}$ matrix with comparatively bigger $Y^{11}_{\nu}$ 
required to fit the neutrino oscillation data subjected to the choice of $M^{11}$.
\subsection{Analysis and future reach }
\label{sec:coll_ans}
We have simulated pair production of RH stop quarks and its subsequent decays as already discussed for the two above mentioned benchmark 
points. The events were generated at the parton level using MadGraph v2.5.5 \cite{Alwall:2011uj,Alwall:2014hca} and subsequently passed 
through Pythia8 \cite{Sjostrand:2014zea} for 
decay, showering and hadronization. We have used nn23lo1 parton distribution function \cite{Ball:2012cx,Ball:2014uwa}. Detector 
simulation was implemented via Delphes v3.4.1 \cite{deFavereau:2013fsa,Selvaggi:2014mya,Mertens:2015kba}. 
Jets were constructed at this stage using anti-kT algorithm \cite{Cacciari:2008gp}. The ATLAS 
Collaboration has recently included the same-sign trilepton channel in their search for stop quarks in the context of 
MSSM \cite{Aaboud:2017dmy}. We have implemented the same analysis to obtain our signal rate. The stop production cross-section 
has been appropriately scaled using the next-to-leading-order (NLO) factor obtained from the LHC SUSY cross-section working group webpage \cite{csec:stop}.

The electrons are primarily selected with $p_T > 10$ GeV and $|\eta| < 2.47$ barring the region between the barrel and endcap electromagnetic 
calorimeters, $1.37 < |\eta| < 1.52$. The muon candidates are selected with the same $p_T$ threshold and $|\eta| < 2.5$. Jets are reconstructed 
with radius $R = 0.4$, $p_T > 20$ GeV and $|\eta| < 2.8$. In our analysis we have implemented the $p_T$ dependent b-jet tagging efficiency and 
light jet misidentification efficiency 
following the ATLAS Collaboration criteria \cite{Aaboud:2017dmy}. Finally, the b-jets are counted with $|\eta| < 2.5$. The event selection further requires 
at least two of the three same-sign leptons have $p_T > 20$ GeV. In order to reduce SM backgrounds arising from mismeasurement of electron charge, 
events are vetoed if invariant mass of two same-sign electrons is within a 10 GeV window of the $Z$-boson mass. The ATLAS Collaboration has obtained 
$1.6\pm 0.8$ SM background events at 13 TeV LHC with these criteria  in the $\ell^{\pm}\ell^{\pm}\ell^{\pm} + \ge$ 1 b-jet signal region with an integrated 
luminosity of 36 $\ifb$  \cite{Aaboud:2017dmy}. In Table~\ref{tab:bp_res} 
below, we have shown the signal cross-section  corresponding to our benchmark points and also the required luminosity to probe this scenario with 
3$\sigma$ and 5$\sigma$ statistical significance ($\mathcal{S}$). In order to compute $\mathcal{S}$ of our signal ($S$)
over the SM background ($B$) we have used $\mathcal{S} = \frac{S}{\sqrt{B + \sigma^2_B}}$,
where $\sigma_B$ is the uncertainty in the measurement of the SM 
background. For simplicity, even for the higher luminosity estimates, we have assumed that $\sigma_B$ remains the same
fraction of $B$ as it is in $36~\ifb$ integrated luminosity\footnote{This is a forced choice we had to make due to lack of information resulting in a conservative
estimate.}. Note that, the IH scenario for BP1 can already be ruled out from the accumulated data. However, the NH case needs an integrated luminosity 
$\sim$ 828.3 $\ifb$ to achieve a $3\sigma$ statistical significance. BP2, on the other hand, would require much higher luminosity owing to the larger stop 
mass. A discovery significance in the NH case is beyond the reach of even 3000 $\ifb$ integrated luminosity. However, for the IH case, 
${\mathcal S} = 3\sigma$ can be achieved at $\sim$ 845.8 $\ifb$.   
\begin{table}[h!]
\begin{center}
\begin{tabular}{||c||c|c||c|c||}
\hline
\multicolumn{1}{||c||}{Results } &
\multicolumn{2}{|c||}{\bf BP1} &
\multicolumn{2}{|c||}{\bf BP2}\\
\cline{2-5}
 & Normal & Inverted & Normal & Inverted \\
\hline\hline
$\sigma_{\rm sig}$ (fb) & 0.070 & 0.193 & 0.026 & 0.070 \\
\hline
Required $\mathcal{L}~(\ifb)~(3\sigma)$ & 828.3 & 12.2 & $> 3000$ & 845.8 \\
Required $\mathcal{L}~(\ifb)~(5\sigma)$ & $> 3000$ & 44.6 & $> 3000$ & $> 3000$ \\
\hline\hline
\end{tabular}
\caption{Signal cross-section and required integrated luminosity at 13 TeV LHC to probe the benchmark points with 3$\sigma$ and 5$\sigma$ 
statistical significance in the $\ell^{\pm}\ell^{\pm}\ell^{\pm} + \ge$ 1 b-jet signal region.}
\label{tab:bp_res}
\end{center}
\end{table}

As evident from Table \ref{tab:bp_res}, the stop mass probe in the present scenario depends heavily on the choice of neutrino mass hierarchy 
due to the variation in BR$(\chonepm\to\ell\lsp)$. For 
more comprehensive understanding of the parameter space, we have used the existing LHC data in order to determine the 
exclusion limit on the lighter stop mass subjected to NH and IH in the light neutrino masses.
For this, we have kept the relative separation of the masses uniform throughout, i.e., $m_{\lsptwo}\le m_{\tilde t_1} - m_t$ 
and $m_h \ge m_{\lsptwo} - m_{\chonepm} (m_{\lspone}) \ge m_W$. The simple requirement of our signal region makes sure that the cuts efficiency 
does not vary significantly with such choice of the spectrum over a wide range of stop mass.   
\begin{figure}[h!]
\includegraphics[scale=0.8]{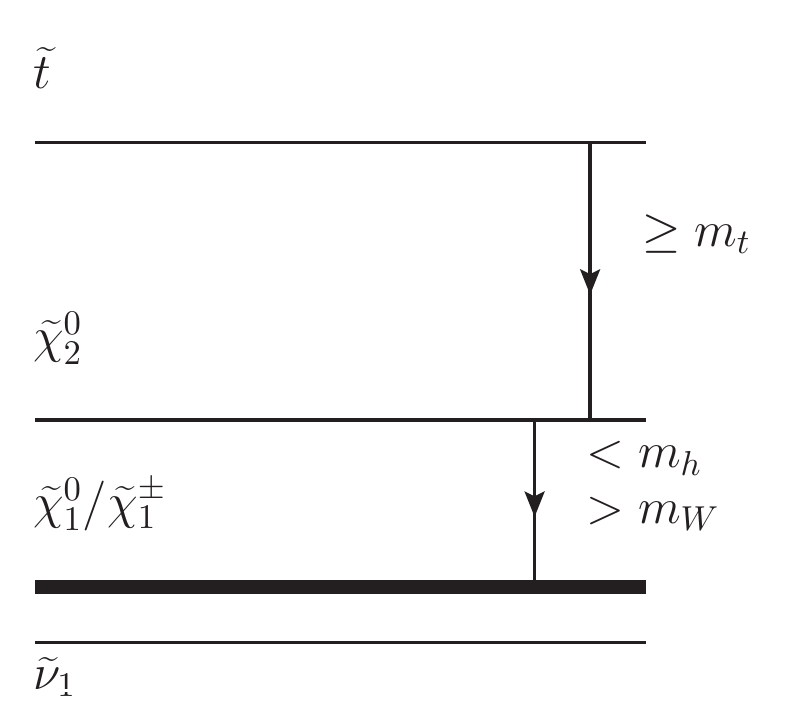}
\includegraphics[scale=0.4]{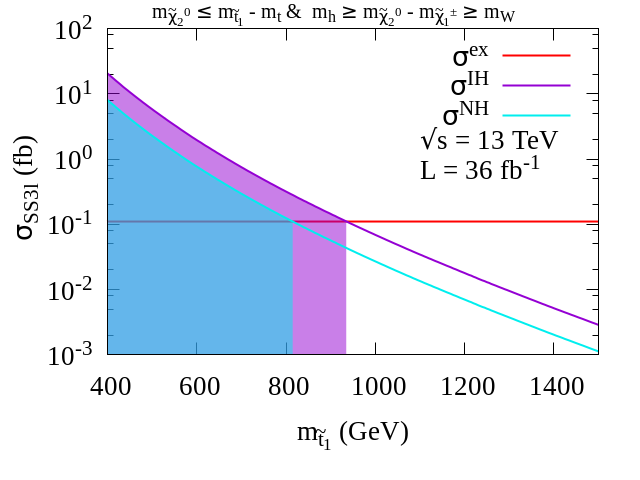}
\caption{On the left: Choice of mass spectra. On the right: The expected cross-section in the same-sign tri-lepton final state vs the stop mass 
and the exclusion obtained from the 36 $\ifb$ luminosity data at the LHC. The red horizontal line corresponds to the experimental limit on the  
visible cross-section \cite{Aaboud:2017dmy}. }
\label{fig:exclu}
\end{figure}

In Fig.~\ref{fig:exclu} we have shown the choice of our mass spectra in the framework of SISM under consideration and presented reach of stop 
masses with existing data at the 13 TeV LHC with 36 $\ifb$ luminosity data. The red horizontal line represents the 95\% confidence level exclusion 
limit on the visible cross-section ($\sigma_{\rm vis.} = 0.11$ fb) in this signal region put by the ATLAS collaboration \cite{Aaboud:2017dmy}. 
The violet and cyan lines represent the 
stop mass reach in our model framework subjected to the existing data with IH and NH choices in the light neutrino masses respectively with the shaded 
regions already excluded. As evident, $m_{\tilde t_1} < 815$ and 935 GeV are already excluded under the NH and IH assumptions respectively. 

If some excess is found in this signal region at higher luminosity, it would be useful to have a prior idea of the mass range in which we can 
hope to discover the stop with certainty. In order to determine this, we have computed the discovery significance of the stop quark lying within the mass 
range [700 GeV - 1.2 TeV] at two different luminosities, $130~\ifb$ and $3000~\ifb$ with $\sqrt{s} = 13$ TeV. Fig.~\ref{fig:fut_excl}
shows the distribution of the obtained statistical significance as a function of $m_{\tilde t_1}$. The horizontal grey lines in the figure 
represent the coveted 3$\sigma$ and 5$\sigma$ statistical significance requirements. The violet and cyan solid lines represent the prospects of our scenario 
subjected to the IH and NH choices. 
The dotted lines around the solid ones are obtained assuming a 10\% variation in $\sigma_B$. 
As evident from Fig.~\ref{fig:fut_excl}, at low luminosity (130 $\ifb$), the discovery reach of $m_{\tilde t_1}$ can be upto $\sim 970$ GeV 
at most whereas, the high luminosity (3000 $\ifb$) option can slightly improve the mass range upto $\sim 1$ TeV. If IH in the neutrino masses is 
ruled out in future by the neutrino oscillation experiments, $m_{\tilde t_1}$ discovery reach can not be beyond 900 GeV even at large luminosity at the LHC.
\begin{figure}[h!]
\centering
\hspace{-1cm}
\includegraphics[scale=0.35]{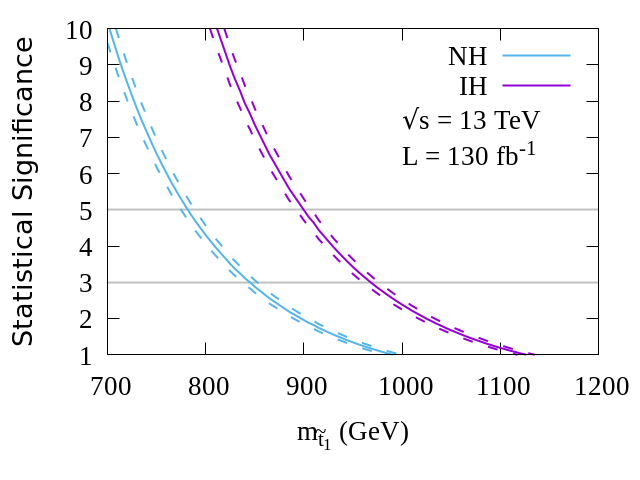}
\includegraphics[scale=0.35]{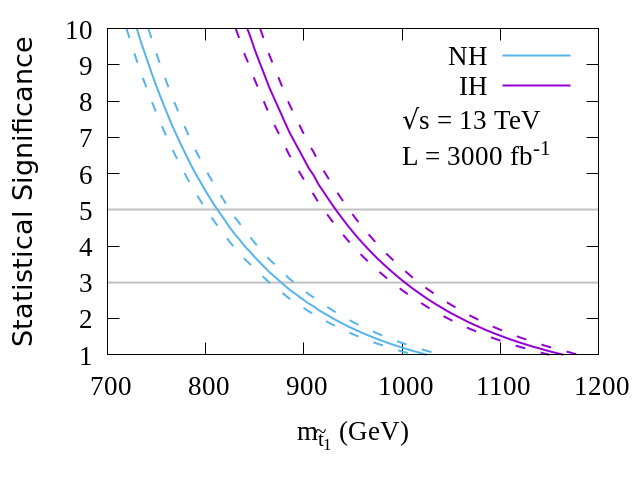}
\caption{Discovery reach in the stop mass plane at $\sqrt{s} = 13$ TeV LHC with two choices of integrated luminosity, 130 $\ifb$ and 3000 $\ifb$. The parallel grey lines 
indicate 3$\sigma$ and 5$\sigma$ statistical significance. }
\label{fig:fut_excl}
\end{figure}

In spite of the small signal cross-section, therefore, the stop mass discovery reach can be in the TeV range in this signal region owing to the small SM 
background. The results do not improve drastically with the increase in the luminosity because of the presence of a large uncertainty factor in the SM background 
estimation. With more accumulation of data the statistical error is expected to decrease and that can improve the stop discovery reach significantly. 
With a high center-of-mass energy hadron collider becoming more and more relevant in the context of SUSY search, we have also estimated a 
projected discovery reach of the stop mass in the present scenario at a 27 TeV pp collider. Our SM background computation at this centre-of-mass energy 
from $t\bar th$, $t\bar t W$, $t\bar tZ(\gamma)$, $VV$ and $VVV$, where $V$ represents gauge bosons ($W$ and $Z$) production channels yields a cross-section 
of 0.077 fb. We have computed $\sigma_B$ following the same prescription as in Fig.~\ref{fig:fut_excl}. The NLO k-factor has been computed with Prospino 
\cite{Beenakker:1996ed,Beenakker:1997ut} The $\sqrt{s} = 27$ TeV results are 
shown in Fig.~\ref{fig:fut_excl_27T}. The increase in the signal cross-section improves the discovery reach of $m_{\widetilde t_1}$  $\sim$ 1.55 TeV 
and $\sim$ 1.45 TeV for IH and NH choices respectively. 
\begin{figure}[h!]
\centering
\hspace{-1cm}
\includegraphics[scale=0.35]{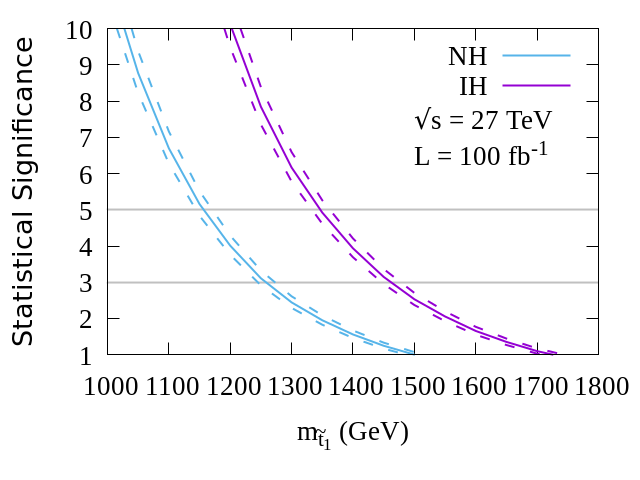}
\includegraphics[scale=0.35]{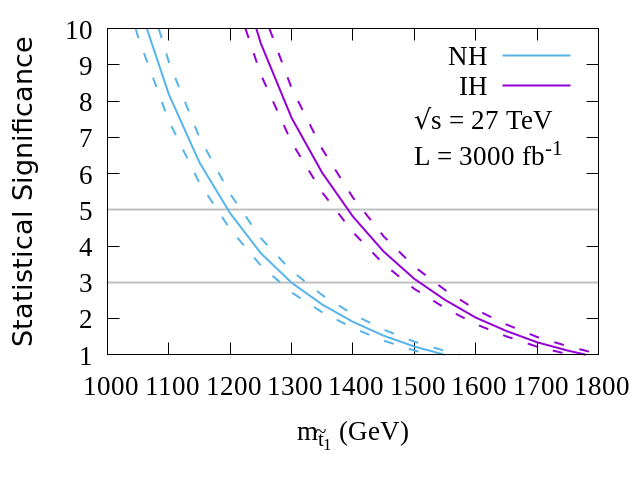}
\caption{Discovery reach in the stop mass plane at $\sqrt{s} = 27$ TeV LHC with two choices of integrated luminosity, 100 $\ifb$ and 3000 $\ifb$. 
Color codings are similar to Fig.~\ref{fig:fut_excl}.}
\label{fig:fut_excl_27T}
\end{figure}

\section{Summary and Conclusion}
\label{sec:concl}
In this work, we have investigated the interplay among neutrino mass mechanism, DM requirements and stop search at the LHC in 
the context of supersymmetric inverse seesaw scenario. The SISM scenario naturally gives rise to a right-sneutrino DM in the 
theory in the form of the LSP within an R-parity conserving framework. Originating from a gauge singlet, the sneutrino requires 
to lie close to the 125 GeV Higgs resonance annihilation in order to produce correct relic density. However, 
the present DM direct detection constraints have put this parameter space in jeopardy forcing us to look for co-annihilation 
of the right-sneutrino DM. 

The sneutrino interactions are driven by the same Dirac neutrino Yukawa couplings introduced to 
fit the neutrino oscillation data. We have shown, using Casas-Ibarra parameterization, that the lepton flavor violating 
constraints force these Yukawa parameters to be small. The choice of $yY_{\nu}$ here is subjected to the choice of heavy neutrino mass 
and the lepton number violating parameter, $\mu_S$. Thus the LFV constraints put a limit on the choices of $\mu_S \gtrsim 10^{-6}$ GeV 
for sub-TeV heavy neutrino masses in particular for IH. 
Among the constraints arising from the decay, $\mu\to e\gamma$ turns out to be the most severe. The predictions for 
$\tau \to \mu \gamma$, and $\tau \to \mu \mu \mu$ are relatively weaker, and constrain sneutrino mass only in the 
$m_{\tilde{\nu}} \lsim 100$ GeV range. The model predictions in the $\tau$ sector can be tested in the future  LFV searches.
The constrained $Y_{\nu}$ from the LFV data results in a very small direct detection cross-section of the sneutrino, which is beyond 
the present DM experimental sensitivity by a few orders of magnitude. This also forces the resonant contribution in relic density to 
be small and motivates one to look for co-annihilation of the 
sneutrino DM with other sparticles. In this work, we have explored the sneutrino-wino co-annihilation which has interesting implications 
for stop search at the LHC. 

The co-annihilating DM scenario gives rise to a partially compressed electroweak spectra which makes the conventional stop search strategies 
weaker. We have shown that the relic density is satisfied due to  co-annihilation of sneutrino with wino like neutralino, and chargino, 
requiring the mass difference between the LSP and chargino/neutralino to be $\Delta m \sim 15-30$ GeV for all values of sneutrino masses 
in between 100-1000 GeV. Indirect limit on the LSP sneutrino mass arises from the wino mass bounds. 
However, the compression in the spectra again results in relatively weaker exclusions allowing the LSP to be as light as 200 GeV in 
this scenario.

Under such circumstances, conventional stop search strategies prove to be ineffective in probing TeV scale stop masses. 
Here we have studied stop pair production and its subsequent decays  into top and a bino-like NNLSP $\tilde{\chi^0}_2$.
Further decay of the bino into gauge boson and a wino-like chargino $\tilde{\chi}^{\pm}$, and $\tilde{\chi}^{\pm} \to \ell \tilde{\nu}_{1,2}$ 
leads to a novel same-sign trilepton signal $\ell^{\pm} \ell^{\pm} \ell^{\pm}+ \ge 1 b-\rm{jets}+\cancel{E_T} $ for the stop at LHC. 
Unlike the 
conventional stop search channels, this signal region has very small SM background. As a result, despite of the small signal rate, stop 
mass can be probed close to 935 (815) GeV with the existing LHC data for IH (NH) choices in neutrino mass hierarchy. 
Stop mass limits prove to be stronger for the inverted hierarchy in the light neutrino masses due to the greater leptonic branching ratio 
of the chargino resulting in greater abundance of electrons in the final state, which also serves as a distinguishing feature of this model 
from the MSSM. In the MSSM with neutralino LSP, however, this same-sign trilepton signal rate will be much lower. 
A $3\sigma$ discovery significance can be obtained with $m_{\widetilde t_1}\sim$ 1 TeV at high luminosity with 13 TeV center-of-mass energy. 
We have also shown that the 
stop mass discovery reach can go beyond 1.5 TeV at a high-energy pp collider with 27 TeV center-of-mass energy. With improved sensitivity 
in SM background measurement at higher luminosity, the conservative limits we obtained are bound to improve significantly.   
\section{Acknowledgement}
SM and KH acknowledge H2020-MSCA-RICE-2014 grant no. 645722 (NonMinimal Higgs).
SM would like to thank IACS, Kolkata for hospitality during the final phase of this work.
M.M. would like to acknowledge the DST-INSPIRE research grant IFA14-PH-99.
\bibliography{stop_invs}

\begin{thebibliography}{92}%
\makeatletter
\providecommand \@ifxundefined [1]{%
 \@ifx{#1\undefined}
}%
\providecommand \@ifnum [1]{%
 \ifnum #1\expandafter \@firstoftwo
 \else \expandafter \@secondoftwo
 \fi
}%
\providecommand \@ifx [1]{%
 \ifx #1\expandafter \@firstoftwo
 \else \expandafter \@secondoftwo
 \fi
}%
\providecommand \natexlab [1]{#1}%
\providecommand \enquote  [1]{``#1''}%
\providecommand \bibnamefont  [1]{#1}%
\providecommand \bibfnamefont [1]{#1}%
\providecommand \citenamefont [1]{#1}%
\providecommand \href@noop [0]{\@secondoftwo}%
\providecommand \href [0]{\begingroup \@sanitize@url \@href}%
\providecommand \@href[1]{\@@startlink{#1}\@@href}%
\providecommand \@@href[1]{\endgroup#1\@@endlink}%
\providecommand \@sanitize@url [0]{\catcode `\\12\catcode `\$12\catcode
  `\&12\catcode `\#12\catcode `\^12\catcode `\_12\catcode `\%12\relax}%
\providecommand \@@startlink[1]{}%
\providecommand \@@endlink[0]{}%
\providecommand \url  [0]{\begingroup\@sanitize@url \@url }%
\providecommand \@url [1]{\endgroup\@href {#1}{\urlprefix }}%
\providecommand \urlprefix  [0]{URL }%
\providecommand \Eprint [0]{\href }%
\providecommand \doibase [0]{http://dx.doi.org/}%
\providecommand \selectlanguage [0]{\@gobble}%
\providecommand \bibinfo  [0]{\@secondoftwo}%
\providecommand \bibfield  [0]{\@secondoftwo}%
\providecommand \translation [1]{[#1]}%
\providecommand \BibitemOpen [0]{}%
\providecommand \bibitemStop [0]{}%
\providecommand \bibitemNoStop [0]{.\EOS\space}%
\providecommand \EOS [0]{\spacefactor3000\relax}%
\providecommand \BibitemShut  [1]{\csname bibitem#1\endcsname}%
\let\auto@bib@innerbib\@empty
\bibitem [{\citenamefont {Gonzalez-Garcia}\ \emph {et~al.}(2016)\citenamefont
  {Gonzalez-Garcia}, \citenamefont {Maltoni},\ and\ \citenamefont
  {Schwetz}}]{Gonzalez-Garcia:2015qrr}%
  \BibitemOpen
  \bibfield  {author} {\bibinfo {author} {\bibfnamefont {M.~C.}\ \bibnamefont
  {Gonzalez-Garcia}}, \bibinfo {author} {\bibfnamefont {M.}~\bibnamefont
  {Maltoni}}, \ and\ \bibinfo {author} {\bibfnamefont {T.}~\bibnamefont
  {Schwetz}},\ }\href {\doibase 10.1016/j.nuclphysb.2016.02.033} {\bibfield
  {journal} {\bibinfo  {journal} {Nucl. Phys.}\ }\textbf {\bibinfo {volume}
  {B908}},\ \bibinfo {pages} {199} (\bibinfo {year} {2016})},\ \Eprint
  {http://arxiv.org/abs/1512.06856} {arXiv:1512.06856 [hep-ph]} \BibitemShut
  {NoStop}%
\bibitem [{\citenamefont {de~Salas}\ \emph {et~al.}(2017)\citenamefont
  {de~Salas}, \citenamefont {Forero}, \citenamefont {Ternes}, \citenamefont
  {Tortola},\ and\ \citenamefont {Valle}}]{deSalas:2017kay}%
  \BibitemOpen
  \bibfield  {author} {\bibinfo {author} {\bibfnamefont {P.~F.}\ \bibnamefont
  {de~Salas}}, \bibinfo {author} {\bibfnamefont {D.~V.}\ \bibnamefont
  {Forero}}, \bibinfo {author} {\bibfnamefont {C.~A.}\ \bibnamefont {Ternes}},
  \bibinfo {author} {\bibfnamefont {M.}~\bibnamefont {Tortola}}, \ and\
  \bibinfo {author} {\bibfnamefont {J.~W.~F.}\ \bibnamefont {Valle}},\
  }\href@noop {} {\  (\bibinfo {year} {2017})},\ \Eprint
  {http://arxiv.org/abs/1708.01186} {arXiv:1708.01186 [hep-ph]} \BibitemShut
  {NoStop}%
\bibitem [{\citenamefont {Minkowski}(1977)}]{Minkowski:1977sc}%
  \BibitemOpen
  \bibfield  {author} {\bibinfo {author} {\bibfnamefont {P.}~\bibnamefont
  {Minkowski}},\ }\href {\doibase 10.1016/0370-2693(77)90435-X} {\bibfield
  {journal} {\bibinfo  {journal} {Phys. Lett.}\ }\textbf {\bibinfo {volume}
  {B67}},\ \bibinfo {pages} {421} (\bibinfo {year} {1977})}\BibitemShut
  {NoStop}%
\bibitem [{\citenamefont {Yanagida}()}]{Yanagida:1979as}%
  \BibitemOpen
  \bibfield  {author} {\bibinfo {author} {\bibfnamefont {T.}~\bibnamefont
  {Yanagida}},\ }\href@noop {} {\ }\bibinfo {note} {In Proceedings of the
  Workshop on the Baryon Number of the Universe and Unified Theories, Tsukuba,
  Japan, 13-14 Feb 1979}\BibitemShut {NoStop}%
\bibitem [{\citenamefont {Mohapatra}\ and\ \citenamefont
  {Senjanovic}(1980)}]{Mohapatra:1979ia}%
  \BibitemOpen
  \bibfield  {author} {\bibinfo {author} {\bibfnamefont {R.~N.}\ \bibnamefont
  {Mohapatra}}\ and\ \bibinfo {author} {\bibfnamefont {G.}~\bibnamefont
  {Senjanovic}},\ }\href {\doibase 10.1103/PhysRevLett.44.912} {\bibfield
  {journal} {\bibinfo  {journal} {Phys. Rev. Lett.}\ }\textbf {\bibinfo
  {volume} {44}},\ \bibinfo {pages} {912} (\bibinfo {year} {1980})}\BibitemShut
  {NoStop}%
\bibitem [{\citenamefont {Glashow}(1980)}]{Glashow:1979nm}%
  \BibitemOpen
  \bibfield  {author} {\bibinfo {author} {\bibfnamefont {S.~L.}\ \bibnamefont
  {Glashow}},\ }\href@noop {} {\bibfield  {journal} {\bibinfo  {journal} {NATO
  Adv. Study Inst. Ser. B Phys.}\ }\textbf {\bibinfo {volume} {59}},\ \bibinfo
  {pages} {687} (\bibinfo {year} {1980})}\BibitemShut {NoStop}%
\bibitem [{\citenamefont {Gell-Mann}\ \emph {et~al.}()\citenamefont
  {Gell-Mann}, \citenamefont {Ramond},\ and\ \citenamefont
  {Slansky}}]{GellMann:1980vs}%
  \BibitemOpen
  \bibfield  {author} {\bibinfo {author} {\bibfnamefont {M.}~\bibnamefont
  {Gell-Mann}}, \bibinfo {author} {\bibfnamefont {P.}~\bibnamefont {Ramond}}, \
  and\ \bibinfo {author} {\bibfnamefont {R.}~\bibnamefont {Slansky}},\
  }\href@noop {} {\ }\bibinfo {note} {Print-80-0576 (CERN)}\BibitemShut
  {NoStop}%
\bibitem [{\citenamefont {Mohapatra}(1986)}]{Mohapatra:1986aw}%
  \BibitemOpen
  \bibfield  {author} {\bibinfo {author} {\bibfnamefont {R.~N.}\ \bibnamefont
  {Mohapatra}},\ }\href {\doibase 10.1103/PhysRevLett.56.561} {\bibfield
  {journal} {\bibinfo  {journal} {Phys. Rev. Lett.}\ }\textbf {\bibinfo
  {volume} {56}},\ \bibinfo {pages} {561} (\bibinfo {year} {1986})}\BibitemShut
  {NoStop}%
\bibitem [{\citenamefont {Mohapatra}\ and\ \citenamefont
  {Valle}(1986)}]{Mohapatra:1986bd}%
  \BibitemOpen
  \bibfield  {author} {\bibinfo {author} {\bibfnamefont {R.~N.}\ \bibnamefont
  {Mohapatra}}\ and\ \bibinfo {author} {\bibfnamefont {J.~W.~F.}\ \bibnamefont
  {Valle}},\ }\bibfield  {booktitle} {\emph {\bibinfo {booktitle} {{Sixty years
  of double beta decay: From nuclear physics to beyond standard model particle
  physics}}},\ }\href {\doibase 10.1103/PhysRevD.34.1642} {\bibfield  {journal}
  {\bibinfo  {journal} {Phys. Rev.}\ }\textbf {\bibinfo {volume} {D34}},\
  \bibinfo {pages} {1642} (\bibinfo {year} {1986})},\ \bibinfo {note}
  {[,235(1986)]}\BibitemShut {NoStop}%
\bibitem [{\citenamefont {Gonzalez-Garcia}\ and\ \citenamefont
  {Valle}(1989)}]{GonzalezGarcia:1988rw}%
  \BibitemOpen
  \bibfield  {author} {\bibinfo {author} {\bibfnamefont {M.~C.}\ \bibnamefont
  {Gonzalez-Garcia}}\ and\ \bibinfo {author} {\bibfnamefont {J.~W.~F.}\
  \bibnamefont {Valle}},\ }\href {\doibase 10.1016/0370-2693(89)91131-3}
  {\bibfield  {journal} {\bibinfo  {journal} {Phys. Lett.}\ }\textbf {\bibinfo
  {volume} {B216}},\ \bibinfo {pages} {360} (\bibinfo {year}
  {1989})}\BibitemShut {NoStop}%
\bibitem [{\citenamefont {Das}\ and\ \citenamefont {Okada}(2013)}]{Das:2012ze}%
  \BibitemOpen
  \bibfield  {author} {\bibinfo {author} {\bibfnamefont {A.}~\bibnamefont
  {Das}}\ and\ \bibinfo {author} {\bibfnamefont {N.}~\bibnamefont {Okada}},\
  }\href {\doibase 10.1103/PhysRevD.88.113001} {\bibfield  {journal} {\bibinfo
  {journal} {Phys. Rev.}\ }\textbf {\bibinfo {volume} {D88}},\ \bibinfo {pages}
  {113001} (\bibinfo {year} {2013})},\ \Eprint {http://arxiv.org/abs/1207.3734}
  {arXiv:1207.3734 [hep-ph]} \BibitemShut {NoStop}%
\bibitem [{\citenamefont {Das}\ and\ \citenamefont
  {Okada}(2016)}]{Das:2015toa}%
  \BibitemOpen
  \bibfield  {author} {\bibinfo {author} {\bibfnamefont {A.}~\bibnamefont
  {Das}}\ and\ \bibinfo {author} {\bibfnamefont {N.}~\bibnamefont {Okada}},\
  }\href {\doibase 10.1103/PhysRevD.93.033003} {\bibfield  {journal} {\bibinfo
  {journal} {Phys. Rev.}\ }\textbf {\bibinfo {volume} {D93}},\ \bibinfo {pages}
  {033003} (\bibinfo {year} {2016})},\ \Eprint
  {http://arxiv.org/abs/1510.04790} {arXiv:1510.04790 [hep-ph]} \BibitemShut
  {NoStop}%
\bibitem [{\citenamefont {Das}\ and\ \citenamefont
  {Okada}(2017)}]{Das:2017nvm}%
  \BibitemOpen
  \bibfield  {author} {\bibinfo {author} {\bibfnamefont {A.}~\bibnamefont
  {Das}}\ and\ \bibinfo {author} {\bibfnamefont {N.}~\bibnamefont {Okada}},\
  }\href {\doibase 10.1016/j.physletb.2017.09.042} {\bibfield  {journal}
  {\bibinfo  {journal} {Phys. Lett.}\ }\textbf {\bibinfo {volume} {B774}},\
  \bibinfo {pages} {32} (\bibinfo {year} {2017})},\ \Eprint
  {http://arxiv.org/abs/1702.04668} {arXiv:1702.04668 [hep-ph]} \BibitemShut
  {NoStop}%
\bibitem [{\citenamefont {Lee}\ \emph {et~al.}(2007)\citenamefont {Lee},
  \citenamefont {Matchev},\ and\ \citenamefont {Nasri}}]{Lee:2007mt}%
  \BibitemOpen
  \bibfield  {author} {\bibinfo {author} {\bibfnamefont {H.-S.}\ \bibnamefont
  {Lee}}, \bibinfo {author} {\bibfnamefont {K.~T.}\ \bibnamefont {Matchev}}, \
  and\ \bibinfo {author} {\bibfnamefont {S.}~\bibnamefont {Nasri}},\ }\href
  {\doibase 10.1103/PhysRevD.76.041302} {\bibfield  {journal} {\bibinfo
  {journal} {Phys. Rev.}\ }\textbf {\bibinfo {volume} {D76}},\ \bibinfo {pages}
  {041302} (\bibinfo {year} {2007})},\ \Eprint
  {http://arxiv.org/abs/hep-ph/0702223} {arXiv:hep-ph/0702223 [HEP-PH]}
  \BibitemShut {NoStop}%
\bibitem [{\citenamefont {Cerdeno}\ \emph {et~al.}(2009)\citenamefont
  {Cerdeno}, \citenamefont {Munoz},\ and\ \citenamefont
  {Seto}}]{Cerdeno:2008ep}%
  \BibitemOpen
  \bibfield  {author} {\bibinfo {author} {\bibfnamefont {D.~G.}\ \bibnamefont
  {Cerdeno}}, \bibinfo {author} {\bibfnamefont {C.}~\bibnamefont {Munoz}}, \
  and\ \bibinfo {author} {\bibfnamefont {O.}~\bibnamefont {Seto}},\ }\href
  {\doibase 10.1103/PhysRevD.79.023510} {\bibfield  {journal} {\bibinfo
  {journal} {Phys. Rev.}\ }\textbf {\bibinfo {volume} {D79}},\ \bibinfo {pages}
  {023510} (\bibinfo {year} {2009})},\ \Eprint {http://arxiv.org/abs/0807.3029}
  {arXiv:0807.3029 [hep-ph]} \BibitemShut {NoStop}%
\bibitem [{\citenamefont {Mondal}\ \emph {et~al.}(2012)\citenamefont {Mondal},
  \citenamefont {Biswas}, \citenamefont {Ghosh},\ and\ \citenamefont
  {Roy}}]{Mondal:2012jv}%
  \BibitemOpen
  \bibfield  {author} {\bibinfo {author} {\bibfnamefont {S.}~\bibnamefont
  {Mondal}}, \bibinfo {author} {\bibfnamefont {S.}~\bibnamefont {Biswas}},
  \bibinfo {author} {\bibfnamefont {P.}~\bibnamefont {Ghosh}}, \ and\ \bibinfo
  {author} {\bibfnamefont {S.}~\bibnamefont {Roy}},\ }\href {\doibase
  10.1007/JHEP05(2012)134} {\bibfield  {journal} {\bibinfo  {journal} {JHEP}\
  }\textbf {\bibinfo {volume} {05}},\ \bibinfo {pages} {134} (\bibinfo {year}
  {2012})},\ \Eprint {http://arxiv.org/abs/1201.1556} {arXiv:1201.1556
  [hep-ph]} \BibitemShut {NoStop}%
\bibitem [{\citenamefont {Bhupal~Dev}\ \emph {et~al.}(2012)\citenamefont
  {Bhupal~Dev}, \citenamefont {Mondal}, \citenamefont {Mukhopadhyaya},\ and\
  \citenamefont {Roy}}]{BhupalDev:2012ru}%
  \BibitemOpen
  \bibfield  {author} {\bibinfo {author} {\bibfnamefont {P.~S.}\ \bibnamefont
  {Bhupal~Dev}}, \bibinfo {author} {\bibfnamefont {S.}~\bibnamefont {Mondal}},
  \bibinfo {author} {\bibfnamefont {B.}~\bibnamefont {Mukhopadhyaya}}, \ and\
  \bibinfo {author} {\bibfnamefont {S.}~\bibnamefont {Roy}},\ }\href {\doibase
  10.1007/JHEP09(2012)110} {\bibfield  {journal} {\bibinfo  {journal} {JHEP}\
  }\textbf {\bibinfo {volume} {09}},\ \bibinfo {pages} {110} (\bibinfo {year}
  {2012})},\ \Eprint {http://arxiv.org/abs/1207.6542} {arXiv:1207.6542
  [hep-ph]} \BibitemShut {NoStop}%
\bibitem [{\citenamefont {De~Romeri}\ and\ \citenamefont
  {Hirsch}(2012)}]{DeRomeri:2012qd}%
  \BibitemOpen
  \bibfield  {author} {\bibinfo {author} {\bibfnamefont {V.}~\bibnamefont
  {De~Romeri}}\ and\ \bibinfo {author} {\bibfnamefont {M.}~\bibnamefont
  {Hirsch}},\ }\href {\doibase 10.1007/JHEP12(2012)106} {\bibfield  {journal}
  {\bibinfo  {journal} {JHEP}\ }\textbf {\bibinfo {volume} {12}},\ \bibinfo
  {pages} {106} (\bibinfo {year} {2012})},\ \Eprint
  {http://arxiv.org/abs/1209.3891} {arXiv:1209.3891 [hep-ph]} \BibitemShut
  {NoStop}%
\bibitem [{\citenamefont {Banerjee}\ \emph {et~al.}(2013)\citenamefont
  {Banerjee}, \citenamefont {Dev}, \citenamefont {Mondal}, \citenamefont
  {Mukhopadhyaya},\ and\ \citenamefont {Roy}}]{Banerjee:2013fga}%
  \BibitemOpen
  \bibfield  {author} {\bibinfo {author} {\bibfnamefont {S.}~\bibnamefont
  {Banerjee}}, \bibinfo {author} {\bibfnamefont {P.~S.~B.}\ \bibnamefont
  {Dev}}, \bibinfo {author} {\bibfnamefont {S.}~\bibnamefont {Mondal}},
  \bibinfo {author} {\bibfnamefont {B.}~\bibnamefont {Mukhopadhyaya}}, \ and\
  \bibinfo {author} {\bibfnamefont {S.}~\bibnamefont {Roy}},\ }\href {\doibase
  10.1007/JHEP10(2013)221} {\bibfield  {journal} {\bibinfo  {journal} {JHEP}\
  }\textbf {\bibinfo {volume} {10}},\ \bibinfo {pages} {221} (\bibinfo {year}
  {2013})},\ \Eprint {http://arxiv.org/abs/1306.2143} {arXiv:1306.2143
  [hep-ph]} \BibitemShut {NoStop}%
\bibitem [{\citenamefont {Ghosh}\ \emph {et~al.}(2015)\citenamefont {Ghosh},
  \citenamefont {Mondal},\ and\ \citenamefont {Saha}}]{Ghosh:2014pwa}%
  \BibitemOpen
  \bibfield  {author} {\bibinfo {author} {\bibfnamefont {D.~K.}\ \bibnamefont
  {Ghosh}}, \bibinfo {author} {\bibfnamefont {S.}~\bibnamefont {Mondal}}, \
  and\ \bibinfo {author} {\bibfnamefont {I.}~\bibnamefont {Saha}},\ }\href
  {\doibase 10.1088/1475-7516/2015/02/035} {\bibfield  {journal} {\bibinfo
  {journal} {JCAP}\ }\textbf {\bibinfo {volume} {1502}},\ \bibinfo {pages}
  {035} (\bibinfo {year} {2015})},\ \Eprint {http://arxiv.org/abs/1405.0206}
  {arXiv:1405.0206 [hep-ph]} \BibitemShut {NoStop}%
\bibitem [{\citenamefont {Chang}\ \emph {et~al.}(2017)\citenamefont {Chang},
  \citenamefont {Cheung}, \citenamefont {Ishida}, \citenamefont {Lu},
  \citenamefont {Spinrath},\ and\ \citenamefont {Tsai}}]{Chang:2017qgi}%
  \BibitemOpen
  \bibfield  {author} {\bibinfo {author} {\bibfnamefont {J.}~\bibnamefont
  {Chang}}, \bibinfo {author} {\bibfnamefont {K.}~\bibnamefont {Cheung}},
  \bibinfo {author} {\bibfnamefont {H.}~\bibnamefont {Ishida}}, \bibinfo
  {author} {\bibfnamefont {C.-T.}\ \bibnamefont {Lu}}, \bibinfo {author}
  {\bibfnamefont {M.}~\bibnamefont {Spinrath}}, \ and\ \bibinfo {author}
  {\bibfnamefont {Y.-L.~S.}\ \bibnamefont {Tsai}},\ }\href {\doibase
  10.1007/JHEP10(2017)039} {\bibfield  {journal} {\bibinfo  {journal} {JHEP}\
  }\textbf {\bibinfo {volume} {10}},\ \bibinfo {pages} {039} (\bibinfo {year}
  {2017})},\ \Eprint {http://arxiv.org/abs/1707.04374} {arXiv:1707.04374
  [hep-ph]} \BibitemShut {NoStop}%
\bibitem [{\citenamefont {Chang}\ \emph {et~al.}(2018)\citenamefont {Chang},
  \citenamefont {Cheung}, \citenamefont {Ishida}, \citenamefont {Lu},
  \citenamefont {Spinrath},\ and\ \citenamefont {Tsai}}]{Chang:2018agk}%
  \BibitemOpen
  \bibfield  {author} {\bibinfo {author} {\bibfnamefont {J.}~\bibnamefont
  {Chang}}, \bibinfo {author} {\bibfnamefont {K.}~\bibnamefont {Cheung}},
  \bibinfo {author} {\bibfnamefont {H.}~\bibnamefont {Ishida}}, \bibinfo
  {author} {\bibfnamefont {C.-T.}\ \bibnamefont {Lu}}, \bibinfo {author}
  {\bibfnamefont {M.}~\bibnamefont {Spinrath}}, \ and\ \bibinfo {author}
  {\bibfnamefont {Y.-L.~S.}\ \bibnamefont {Tsai}},\ }\href@noop {} {\
  (\bibinfo {year} {2018})},\ \Eprint {http://arxiv.org/abs/1806.04468}
  {arXiv:1806.04468 [hep-ph]} \BibitemShut {NoStop}%
\bibitem [{\citenamefont {Delle~Rose}\ \emph {et~al.}(2017)\citenamefont
  {Delle~Rose}, \citenamefont {Khalil}, \citenamefont {King}, \citenamefont
  {Marzo}, \citenamefont {Moretti},\ and\ \citenamefont
  {Un}}]{DelleRose:2017ukx}%
  \BibitemOpen
  \bibfield  {author} {\bibinfo {author} {\bibfnamefont {L.}~\bibnamefont
  {Delle~Rose}}, \bibinfo {author} {\bibfnamefont {S.}~\bibnamefont {Khalil}},
  \bibinfo {author} {\bibfnamefont {S.~J.~D.}\ \bibnamefont {King}}, \bibinfo
  {author} {\bibfnamefont {C.}~\bibnamefont {Marzo}}, \bibinfo {author}
  {\bibfnamefont {S.}~\bibnamefont {Moretti}}, \ and\ \bibinfo {author}
  {\bibfnamefont {C.~S.}\ \bibnamefont {Un}},\ }\href {\doibase
  10.1103/PhysRevD.96.055004} {\bibfield  {journal} {\bibinfo  {journal} {Phys.
  Rev.}\ }\textbf {\bibinfo {volume} {D96}},\ \bibinfo {pages} {055004}
  (\bibinfo {year} {2017})},\ \Eprint {http://arxiv.org/abs/1702.01808}
  {arXiv:1702.01808 [hep-ph]} \BibitemShut {NoStop}%
\bibitem [{\citenamefont {Delle~Rose}\ \emph
  {et~al.}(2018{\natexlab{a}})\citenamefont {Delle~Rose}, \citenamefont
  {Khalil}, \citenamefont {King}, \citenamefont {Kulkarni}, \citenamefont
  {Marzo}, \citenamefont {Moretti},\ and\ \citenamefont
  {Un}}]{DelleRose:2017uas}%
  \BibitemOpen
  \bibfield  {author} {\bibinfo {author} {\bibfnamefont {L.}~\bibnamefont
  {Delle~Rose}}, \bibinfo {author} {\bibfnamefont {S.}~\bibnamefont {Khalil}},
  \bibinfo {author} {\bibfnamefont {S.~J.~D.}\ \bibnamefont {King}}, \bibinfo
  {author} {\bibfnamefont {S.}~\bibnamefont {Kulkarni}}, \bibinfo {author}
  {\bibfnamefont {C.}~\bibnamefont {Marzo}}, \bibinfo {author} {\bibfnamefont
  {S.}~\bibnamefont {Moretti}}, \ and\ \bibinfo {author} {\bibfnamefont
  {C.~S.}\ \bibnamefont {Un}},\ }\href {\doibase 10.1007/JHEP07(2018)100}
  {\bibfield  {journal} {\bibinfo  {journal} {JHEP}\ }\textbf {\bibinfo
  {volume} {07}},\ \bibinfo {pages} {100} (\bibinfo {year}
  {2018}{\natexlab{a}})},\ \Eprint {http://arxiv.org/abs/1712.05232}
  {arXiv:1712.05232 [hep-ph]} \BibitemShut {NoStop}%
\bibitem [{\citenamefont {Delle~Rose}\ \emph
  {et~al.}(2018{\natexlab{b}})\citenamefont {Delle~Rose}, \citenamefont
  {Khalil}, \citenamefont {King}, \citenamefont {Kulkarni}, \citenamefont
  {Marzo}, \citenamefont {Moretti},\ and\ \citenamefont
  {Un}}]{DelleRose:2018mjj}%
  \BibitemOpen
  \bibfield  {author} {\bibinfo {author} {\bibfnamefont {L.}~\bibnamefont
  {Delle~Rose}}, \bibinfo {author} {\bibfnamefont {S.}~\bibnamefont {Khalil}},
  \bibinfo {author} {\bibfnamefont {S.~J.~D.}\ \bibnamefont {King}}, \bibinfo
  {author} {\bibfnamefont {S.}~\bibnamefont {Kulkarni}}, \bibinfo {author}
  {\bibfnamefont {C.}~\bibnamefont {Marzo}}, \bibinfo {author} {\bibfnamefont
  {S.}~\bibnamefont {Moretti}}, \ and\ \bibinfo {author} {\bibfnamefont
  {C.~S.}\ \bibnamefont {Un}}\ }(\bibinfo {year} {2018})\ \Eprint
  {http://arxiv.org/abs/1804.09470} {arXiv:1804.09470 [hep-ph]} \BibitemShut
  {NoStop}%
\bibitem [{\citenamefont {Falk}\ \emph {et~al.}(1994)\citenamefont {Falk},
  \citenamefont {Olive},\ and\ \citenamefont {Srednicki}}]{Falk:1994es}%
  \BibitemOpen
  \bibfield  {author} {\bibinfo {author} {\bibfnamefont {T.}~\bibnamefont
  {Falk}}, \bibinfo {author} {\bibfnamefont {K.~A.}\ \bibnamefont {Olive}}, \
  and\ \bibinfo {author} {\bibfnamefont {M.}~\bibnamefont {Srednicki}},\ }\href
  {\doibase 10.1016/0370-2693(94)90639-4} {\bibfield  {journal} {\bibinfo
  {journal} {Phys. Lett.}\ }\textbf {\bibinfo {volume} {B339}},\ \bibinfo
  {pages} {248} (\bibinfo {year} {1994})},\ \Eprint
  {http://arxiv.org/abs/hep-ph/9409270} {arXiv:hep-ph/9409270 [hep-ph]}
  \BibitemShut {NoStop}%
\bibitem [{\citenamefont {Hebbeker}(1999)}]{Hebbeker:1999pi}%
  \BibitemOpen
  \bibfield  {author} {\bibinfo {author} {\bibfnamefont {T.}~\bibnamefont
  {Hebbeker}},\ }\href {\doibase 10.1016/S0370-2693(99)01313-1} {\bibfield
  {journal} {\bibinfo  {journal} {Phys. Lett.}\ }\textbf {\bibinfo {volume}
  {B470}},\ \bibinfo {pages} {259} (\bibinfo {year} {1999})},\ \Eprint
  {http://arxiv.org/abs/hep-ph/9910326} {arXiv:hep-ph/9910326 [hep-ph]}
  \BibitemShut {NoStop}%
\bibitem [{\citenamefont {Aaboud}\ \emph
  {et~al.}(2017{\natexlab{a}})\citenamefont {Aaboud} \emph
  {et~al.}}]{Aaboud:2017dmy}%
  \BibitemOpen
  \bibfield  {author} {\bibinfo {author} {\bibfnamefont {M.}~\bibnamefont
  {Aaboud}} \emph {et~al.} (\bibinfo {collaboration} {ATLAS}),\ }\href
  {\doibase 10.1007/JHEP09(2017)084} {\bibfield  {journal} {\bibinfo  {journal}
  {JHEP}\ }\textbf {\bibinfo {volume} {09}},\ \bibinfo {pages} {084} (\bibinfo
  {year} {2017}{\natexlab{a}})},\ \Eprint {http://arxiv.org/abs/1706.03731}
  {arXiv:1706.03731 [hep-ex]} \BibitemShut {NoStop}%
\bibitem [{\citenamefont {Aaboud}\ \emph
  {et~al.}(2017{\natexlab{b}})\citenamefont {Aaboud} \emph
  {et~al.}}]{Aaboud:2017ejf}%
  \BibitemOpen
  \bibfield  {author} {\bibinfo {author} {\bibfnamefont {M.}~\bibnamefont
  {Aaboud}} \emph {et~al.} (\bibinfo {collaboration} {ATLAS}),\ }\href
  {\doibase 10.1007/JHEP08(2017)006} {\bibfield  {journal} {\bibinfo  {journal}
  {JHEP}\ }\textbf {\bibinfo {volume} {08}},\ \bibinfo {pages} {006} (\bibinfo
  {year} {2017}{\natexlab{b}})},\ \Eprint {http://arxiv.org/abs/1706.03986}
  {arXiv:1706.03986 [hep-ex]} \BibitemShut {NoStop}%
\bibitem [{\citenamefont {Aaboud}\ \emph
  {et~al.}(2017{\natexlab{c}})\citenamefont {Aaboud} \emph
  {et~al.}}]{Aaboud:2017nfd}%
  \BibitemOpen
  \bibfield  {author} {\bibinfo {author} {\bibfnamefont {M.}~\bibnamefont
  {Aaboud}} \emph {et~al.} (\bibinfo {collaboration} {ATLAS}),\ }\href
  {\doibase 10.1140/epjc/s10052-017-5445-x} {\bibfield  {journal} {\bibinfo
  {journal} {Eur. Phys. J.}\ }\textbf {\bibinfo {volume} {C77}},\ \bibinfo
  {pages} {898} (\bibinfo {year} {2017}{\natexlab{c}})},\ \Eprint
  {http://arxiv.org/abs/1708.03247} {arXiv:1708.03247 [hep-ex]} \BibitemShut
  {NoStop}%
\bibitem [{\citenamefont {Aaboud}\ \emph
  {et~al.}(2017{\natexlab{d}})\citenamefont {Aaboud} \emph
  {et~al.}}]{Aaboud:2017wqg}%
  \BibitemOpen
  \bibfield  {author} {\bibinfo {author} {\bibfnamefont {M.}~\bibnamefont
  {Aaboud}} \emph {et~al.} (\bibinfo {collaboration} {ATLAS}),\ }\href
  {\doibase 10.1007/JHEP11(2017)195} {\bibfield  {journal} {\bibinfo  {journal}
  {JHEP}\ }\textbf {\bibinfo {volume} {11}},\ \bibinfo {pages} {195} (\bibinfo
  {year} {2017}{\natexlab{d}})},\ \Eprint {http://arxiv.org/abs/1708.09266}
  {arXiv:1708.09266 [hep-ex]} \BibitemShut {NoStop}%
\bibitem [{\citenamefont {Aaboud}\ \emph
  {et~al.}(2017{\natexlab{e}})\citenamefont {Aaboud} \emph
  {et~al.}}]{Aaboud:2017ayj}%
  \BibitemOpen
  \bibfield  {author} {\bibinfo {author} {\bibfnamefont {M.}~\bibnamefont
  {Aaboud}} \emph {et~al.} (\bibinfo {collaboration} {ATLAS}),\ }\href
  {\doibase 10.1007/JHEP12(2017)085} {\bibfield  {journal} {\bibinfo  {journal}
  {JHEP}\ }\textbf {\bibinfo {volume} {12}},\ \bibinfo {pages} {085} (\bibinfo
  {year} {2017}{\natexlab{e}})},\ \Eprint {http://arxiv.org/abs/1709.04183}
  {arXiv:1709.04183 [hep-ex]} \BibitemShut {NoStop}%
\bibitem [{\citenamefont {Aaboud}\ \emph
  {et~al.}(2017{\natexlab{f}})\citenamefont {Aaboud} \emph
  {et~al.}}]{Aaboud:2017aeu}%
  \BibitemOpen
  \bibfield  {author} {\bibinfo {author} {\bibfnamefont {M.}~\bibnamefont
  {Aaboud}} \emph {et~al.} (\bibinfo {collaboration} {ATLAS}),\ }\href@noop {}
  {\  (\bibinfo {year} {2017}{\natexlab{f}})},\ \Eprint
  {http://arxiv.org/abs/1711.11520} {arXiv:1711.11520 [hep-ex]} \BibitemShut
  {NoStop}%
\bibitem [{\citenamefont {Aaboud}\ \emph
  {et~al.}(2018{\natexlab{a}})\citenamefont {Aaboud} \emph
  {et~al.}}]{Aaboud:2018kya}%
  \BibitemOpen
  \bibfield  {author} {\bibinfo {author} {\bibfnamefont {M.}~\bibnamefont
  {Aaboud}} \emph {et~al.} (\bibinfo {collaboration} {ATLAS}),\ }\href@noop {}
  {\  (\bibinfo {year} {2018}{\natexlab{a}})},\ \Eprint
  {http://arxiv.org/abs/1803.10178} {arXiv:1803.10178 [hep-ex]} \BibitemShut
  {NoStop}%
\bibitem [{\citenamefont {Deppisch}\ \emph {et~al.}(2015)\citenamefont
  {Deppisch}, \citenamefont {Bhupal~Dev},\ and\ \citenamefont
  {Pilaftsis}}]{Deppisch:2015qwa}%
  \BibitemOpen
  \bibfield  {author} {\bibinfo {author} {\bibfnamefont {F.~F.}\ \bibnamefont
  {Deppisch}}, \bibinfo {author} {\bibfnamefont {P.~S.}\ \bibnamefont
  {Bhupal~Dev}}, \ and\ \bibinfo {author} {\bibfnamefont {A.}~\bibnamefont
  {Pilaftsis}},\ }\href {\doibase 10.1088/1367-2630/17/7/075019} {\bibfield
  {journal} {\bibinfo  {journal} {New J. Phys.}\ }\textbf {\bibinfo {volume}
  {17}},\ \bibinfo {pages} {075019} (\bibinfo {year} {2015})},\ \Eprint
  {http://arxiv.org/abs/1502.06541} {arXiv:1502.06541 [hep-ph]} \BibitemShut
  {NoStop}%
\bibitem [{\citenamefont {Khalil}(2010)}]{Khalil:2010iu}%
  \BibitemOpen
  \bibfield  {author} {\bibinfo {author} {\bibfnamefont {S.}~\bibnamefont
  {Khalil}},\ }\href {\doibase 10.1103/PhysRevD.82.077702} {\bibfield
  {journal} {\bibinfo  {journal} {Phys. Rev.}\ }\textbf {\bibinfo {volume}
  {D82}},\ \bibinfo {pages} {077702} (\bibinfo {year} {2010})},\ \Eprint
  {http://arxiv.org/abs/1004.0013} {arXiv:1004.0013 [hep-ph]} \BibitemShut
  {NoStop}%
\bibitem [{\citenamefont {Dev}\ and\ \citenamefont
  {Mohapatra}(2010)}]{Dev:2009aw}%
  \BibitemOpen
  \bibfield  {author} {\bibinfo {author} {\bibfnamefont {P.~S.~B.}\
  \bibnamefont {Dev}}\ and\ \bibinfo {author} {\bibfnamefont {R.~N.}\
  \bibnamefont {Mohapatra}},\ }\href {\doibase 10.1103/PhysRevD.81.013001}
  {\bibfield  {journal} {\bibinfo  {journal} {Phys. Rev.}\ }\textbf {\bibinfo
  {volume} {D81}},\ \bibinfo {pages} {013001} (\bibinfo {year} {2010})},\
  \Eprint {http://arxiv.org/abs/0910.3924} {arXiv:0910.3924 [hep-ph]}
  \BibitemShut {NoStop}%
\bibitem [{\citenamefont {Deppisch}\ and\ \citenamefont
  {Valle}(2005)}]{Deppisch:2004fa}%
  \BibitemOpen
  \bibfield  {author} {\bibinfo {author} {\bibfnamefont {F.}~\bibnamefont
  {Deppisch}}\ and\ \bibinfo {author} {\bibfnamefont {J.~W.~F.}\ \bibnamefont
  {Valle}},\ }\href {\doibase 10.1103/PhysRevD.72.036001} {\bibfield  {journal}
  {\bibinfo  {journal} {Phys. Rev.}\ }\textbf {\bibinfo {volume} {D72}},\
  \bibinfo {pages} {036001} (\bibinfo {year} {2005})},\ \Eprint
  {http://arxiv.org/abs/hep-ph/0406040} {arXiv:hep-ph/0406040 [hep-ph]}
  \BibitemShut {NoStop}%
\bibitem [{\citenamefont {Hirsch}\ \emph {et~al.}(2010)\citenamefont {Hirsch},
  \citenamefont {Kernreiter}, \citenamefont {Romao},\ and\ \citenamefont
  {Villanova~del Moral}}]{Hirsch:2009ra}%
  \BibitemOpen
  \bibfield  {author} {\bibinfo {author} {\bibfnamefont {M.}~\bibnamefont
  {Hirsch}}, \bibinfo {author} {\bibfnamefont {T.}~\bibnamefont {Kernreiter}},
  \bibinfo {author} {\bibfnamefont {J.~C.}\ \bibnamefont {Romao}}, \ and\
  \bibinfo {author} {\bibfnamefont {A.}~\bibnamefont {Villanova~del Moral}},\
  }\href {\doibase 10.1007/JHEP01(2010)103} {\bibfield  {journal} {\bibinfo
  {journal} {JHEP}\ }\textbf {\bibinfo {volume} {01}},\ \bibinfo {pages} {103}
  (\bibinfo {year} {2010})},\ \Eprint {http://arxiv.org/abs/0910.2435}
  {arXiv:0910.2435 [hep-ph]} \BibitemShut {NoStop}%
\bibitem [{\citenamefont {Abada}\ \emph
  {et~al.}(2012{\natexlab{a}})\citenamefont {Abada}, \citenamefont {Das},
  \citenamefont {Vicente},\ and\ \citenamefont {Weiland}}]{Abada:2012cq}%
  \BibitemOpen
  \bibfield  {author} {\bibinfo {author} {\bibfnamefont {A.}~\bibnamefont
  {Abada}}, \bibinfo {author} {\bibfnamefont {D.}~\bibnamefont {Das}}, \bibinfo
  {author} {\bibfnamefont {A.}~\bibnamefont {Vicente}}, \ and\ \bibinfo
  {author} {\bibfnamefont {C.}~\bibnamefont {Weiland}},\ }\href {\doibase
  10.1007/JHEP09(2012)015} {\bibfield  {journal} {\bibinfo  {journal} {JHEP}\
  }\textbf {\bibinfo {volume} {09}},\ \bibinfo {pages} {015} (\bibinfo {year}
  {2012}{\natexlab{a}})},\ \Eprint {http://arxiv.org/abs/1206.6497}
  {arXiv:1206.6497 [hep-ph]} \BibitemShut {NoStop}%
\bibitem [{\citenamefont {Abada}\ \emph {et~al.}(2014)\citenamefont {Abada},
  \citenamefont {Krauss}, \citenamefont {Porod}, \citenamefont {Staub},
  \citenamefont {Vicente},\ and\ \citenamefont {Weiland}}]{Abada:2014kba}%
  \BibitemOpen
  \bibfield  {author} {\bibinfo {author} {\bibfnamefont {A.}~\bibnamefont
  {Abada}}, \bibinfo {author} {\bibfnamefont {M.~E.}\ \bibnamefont {Krauss}},
  \bibinfo {author} {\bibfnamefont {W.}~\bibnamefont {Porod}}, \bibinfo
  {author} {\bibfnamefont {F.}~\bibnamefont {Staub}}, \bibinfo {author}
  {\bibfnamefont {A.}~\bibnamefont {Vicente}}, \ and\ \bibinfo {author}
  {\bibfnamefont {C.}~\bibnamefont {Weiland}},\ }\href {\doibase
  10.1007/JHEP11(2014)048} {\bibfield  {journal} {\bibinfo  {journal} {JHEP}\
  }\textbf {\bibinfo {volume} {11}},\ \bibinfo {pages} {048} (\bibinfo {year}
  {2014})},\ \Eprint {http://arxiv.org/abs/1408.0138} {arXiv:1408.0138
  [hep-ph]} \BibitemShut {NoStop}%
\bibitem [{\citenamefont {Casas}\ and\ \citenamefont
  {Ibarra}(2001)}]{Casas:2001sr}%
  \BibitemOpen
  \bibfield  {author} {\bibinfo {author} {\bibfnamefont {J.~A.}\ \bibnamefont
  {Casas}}\ and\ \bibinfo {author} {\bibfnamefont {A.}~\bibnamefont {Ibarra}},\
  }\href {\doibase 10.1016/S0550-3213(01)00475-8} {\bibfield  {journal}
  {\bibinfo  {journal} {Nucl. Phys.}\ }\textbf {\bibinfo {volume} {B618}},\
  \bibinfo {pages} {171} (\bibinfo {year} {2001})},\ \Eprint
  {http://arxiv.org/abs/hep-ph/0103065} {arXiv:hep-ph/0103065 [hep-ph]}
  \BibitemShut {NoStop}%
\bibitem [{\citenamefont {Sun}\ \emph {et~al.}(2013)\citenamefont {Sun},
  \citenamefont {Feng}, \citenamefont {Luo}, \citenamefont {Yang},\ and\
  \citenamefont {Chen}}]{Sun:2013kga}%
  \BibitemOpen
  \bibfield  {author} {\bibinfo {author} {\bibfnamefont {K.-S.}\ \bibnamefont
  {Sun}}, \bibinfo {author} {\bibfnamefont {T.-F.}\ \bibnamefont {Feng}},
  \bibinfo {author} {\bibfnamefont {G.-H.}\ \bibnamefont {Luo}}, \bibinfo
  {author} {\bibfnamefont {X.-Y.}\ \bibnamefont {Yang}}, \ and\ \bibinfo
  {author} {\bibfnamefont {J.-B.}\ \bibnamefont {Chen}},\ }\href {\doibase
  10.1142/S0217732313501514} {\bibfield  {journal} {\bibinfo  {journal} {Mod.
  Phys. Lett.}\ }\textbf {\bibinfo {volume} {A28}},\ \bibinfo {pages} {1350151}
  (\bibinfo {year} {2013})},\ \Eprint {http://arxiv.org/abs/1312.2073}
  {arXiv:1312.2073 [hep-ph]} \BibitemShut {NoStop}%
\bibitem [{\citenamefont {Marcano}(2017)}]{Marcano:2017ucg}%
  \BibitemOpen
  \bibfield  {author} {\bibinfo {author} {\bibfnamefont {X.}~\bibnamefont
  {Marcano}},\ }\emph {\bibinfo {title} {{Lepton flavor violation from low
  scale seesaw neutrinos with masses reachable at the LHC}}},\ \href
  {https://inspirehep.net/record/1632013/files/arXiv:1710.08032.pdf} {Ph.D.
  thesis},\ \bibinfo  {school} {Madrid, Autonoma U.} (\bibinfo {year} {2017}),\
  \Eprint {http://arxiv.org/abs/1710.08032} {arXiv:1710.08032 [hep-ph]}
  \BibitemShut {NoStop}%
\bibitem [{\citenamefont {Lello}\ and\ \citenamefont
  {Boyanovsky}(2013)}]{Lello:2012gi}%
  \BibitemOpen
  \bibfield  {author} {\bibinfo {author} {\bibfnamefont {L.}~\bibnamefont
  {Lello}}\ and\ \bibinfo {author} {\bibfnamefont {D.}~\bibnamefont
  {Boyanovsky}},\ }\href {\doibase 10.1103/PhysRevD.87.073017} {\bibfield
  {journal} {\bibinfo  {journal} {Phys. Rev.}\ }\textbf {\bibinfo {volume}
  {D87}},\ \bibinfo {pages} {073017} (\bibinfo {year} {2013})},\ \Eprint
  {http://arxiv.org/abs/1208.5559} {arXiv:1208.5559 [hep-ph]} \BibitemShut
  {NoStop}%
\bibitem [{\citenamefont {Blondel}\ \emph {et~al.}(2016)\citenamefont
  {Blondel}, \citenamefont {Graverini}, \citenamefont {Serra},\ and\
  \citenamefont {Shaposhnikov}}]{Blondel:2014bra}%
  \BibitemOpen
  \bibfield  {author} {\bibinfo {author} {\bibfnamefont {A.}~\bibnamefont
  {Blondel}}, \bibinfo {author} {\bibfnamefont {E.}~\bibnamefont {Graverini}},
  \bibinfo {author} {\bibfnamefont {N.}~\bibnamefont {Serra}}, \ and\ \bibinfo
  {author} {\bibfnamefont {M.}~\bibnamefont {Shaposhnikov}} (\bibinfo
  {collaboration} {FCC-ee study Team}),\ }\bibfield  {booktitle} {\emph
  {\bibinfo {booktitle} {{Proceedings, 37th International Conference on High
  Energy Physics (ICHEP 2014): Valencia, Spain, July 2-9, 2014}}},\ }\href
  {\doibase 10.1016/j.nuclphysbps.2015.09.304} {\bibfield  {journal} {\bibinfo
  {journal} {Nucl. Part. Phys. Proc.}\ }\textbf {\bibinfo {volume} {273-275}},\
  \bibinfo {pages} {1883} (\bibinfo {year} {2016})},\ \Eprint
  {http://arxiv.org/abs/1411.5230} {arXiv:1411.5230 [hep-ex]} \BibitemShut
  {NoStop}%
\bibitem [{\citenamefont {Anelli}\ \emph {et~al.}(2015)\citenamefont {Anelli}
  \emph {et~al.}}]{Anelli:2015pba}%
  \BibitemOpen
  \bibfield  {author} {\bibinfo {author} {\bibfnamefont {M.}~\bibnamefont
  {Anelli}} \emph {et~al.} (\bibinfo {collaboration} {SHiP}),\ }\href@noop {}
  {\  (\bibinfo {year} {2015})},\ \Eprint {http://arxiv.org/abs/1504.04956}
  {arXiv:1504.04956 [physics.ins-det]} \BibitemShut {NoStop}%
\bibitem [{\citenamefont {Abada}\ \emph
  {et~al.}(2012{\natexlab{b}})\citenamefont {Abada}, \citenamefont {Das},\ and\
  \citenamefont {Weiland}}]{Abada:2011hm}%
  \BibitemOpen
  \bibfield  {author} {\bibinfo {author} {\bibfnamefont {A.}~\bibnamefont
  {Abada}}, \bibinfo {author} {\bibfnamefont {D.}~\bibnamefont {Das}}, \ and\
  \bibinfo {author} {\bibfnamefont {C.}~\bibnamefont {Weiland}},\ }\href
  {\doibase 10.1007/JHEP03(2012)100} {\bibfield  {journal} {\bibinfo  {journal}
  {JHEP}\ }\textbf {\bibinfo {volume} {03}},\ \bibinfo {pages} {100} (\bibinfo
  {year} {2012}{\natexlab{b}})},\ \Eprint {http://arxiv.org/abs/1111.5836}
  {arXiv:1111.5836 [hep-ph]} \BibitemShut {NoStop}%
\bibitem [{\citenamefont {Adam}\ \emph {et~al.}(2013)\citenamefont {Adam} \emph
  {et~al.}}]{Adam:2013mnn}%
  \BibitemOpen
  \bibfield  {author} {\bibinfo {author} {\bibfnamefont {J.}~\bibnamefont
  {Adam}} \emph {et~al.} (\bibinfo {collaboration} {MEG}),\ }\href {\doibase
  10.1103/PhysRevLett.110.201801} {\bibfield  {journal} {\bibinfo  {journal}
  {Phys. Rev. Lett.}\ }\textbf {\bibinfo {volume} {110}},\ \bibinfo {pages}
  {201801} (\bibinfo {year} {2013})},\ \Eprint {http://arxiv.org/abs/1303.0754}
  {arXiv:1303.0754 [hep-ex]} \BibitemShut {NoStop}%
\bibitem [{\citenamefont {Baldini}\ \emph {et~al.}(2013)\citenamefont {Baldini}
  \emph {et~al.}}]{Baldini:2013ke}%
  \BibitemOpen
  \bibfield  {author} {\bibinfo {author} {\bibfnamefont {A.~M.}\ \bibnamefont
  {Baldini}} \emph {et~al.},\ }\href@noop {} {\  (\bibinfo {year} {2013})},\
  \Eprint {http://arxiv.org/abs/1301.7225} {arXiv:1301.7225 [physics.ins-det]}
  \BibitemShut {NoStop}%
\bibitem [{\citenamefont {Aubert}\ \emph {et~al.}(2010)\citenamefont {Aubert}
  \emph {et~al.}}]{Aubert:2009ag}%
  \BibitemOpen
  \bibfield  {author} {\bibinfo {author} {\bibfnamefont {B.}~\bibnamefont
  {Aubert}} \emph {et~al.} (\bibinfo {collaboration} {BaBar}),\ }\href
  {\doibase 10.1103/PhysRevLett.104.021802} {\bibfield  {journal} {\bibinfo
  {journal} {Phys. Rev. Lett.}\ }\textbf {\bibinfo {volume} {104}},\ \bibinfo
  {pages} {021802} (\bibinfo {year} {2010})},\ \Eprint
  {http://arxiv.org/abs/0908.2381} {arXiv:0908.2381 [hep-ex]} \BibitemShut
  {NoStop}%
\bibitem [{\citenamefont {Aushev}\ \emph {et~al.}(2010)\citenamefont {Aushev}
  \emph {et~al.}}]{Aushev:2010bq}%
  \BibitemOpen
  \bibfield  {author} {\bibinfo {author} {\bibfnamefont {T.}~\bibnamefont
  {Aushev}} \emph {et~al.},\ }\href@noop {} {\  (\bibinfo {year} {2010})},\
  \Eprint {http://arxiv.org/abs/1002.5012} {arXiv:1002.5012 [hep-ex]}
  \BibitemShut {NoStop}%
\bibitem [{\citenamefont {Hayasaka}\ \emph {et~al.}(2010)\citenamefont
  {Hayasaka} \emph {et~al.}}]{Hayasaka:2010np}%
  \BibitemOpen
  \bibfield  {author} {\bibinfo {author} {\bibfnamefont {K.}~\bibnamefont
  {Hayasaka}} \emph {et~al.},\ }\href {\doibase 10.1016/j.physletb.2010.03.037}
  {\bibfield  {journal} {\bibinfo  {journal} {Phys. Lett.}\ }\textbf {\bibinfo
  {volume} {B687}},\ \bibinfo {pages} {139} (\bibinfo {year} {2010})},\ \Eprint
  {http://arxiv.org/abs/1001.3221} {arXiv:1001.3221 [hep-ex]} \BibitemShut
  {NoStop}%
\bibitem [{\citenamefont {Bertl}\ \emph {et~al.}(2006)\citenamefont {Bertl}
  \emph {et~al.}}]{Bertl:2006up}%
  \BibitemOpen
  \bibfield  {author} {\bibinfo {author} {\bibfnamefont {W.~H.}\ \bibnamefont
  {Bertl}} \emph {et~al.} (\bibinfo {collaboration} {SINDRUM II}),\ }\href
  {\doibase 10.1140/epjc/s2006-02582-x} {\bibfield  {journal} {\bibinfo
  {journal} {Eur. Phys. J.}\ }\textbf {\bibinfo {volume} {C47}},\ \bibinfo
  {pages} {337} (\bibinfo {year} {2006})}\BibitemShut {NoStop}%
\bibitem [{\citenamefont {Bartoszek}\ \emph {et~al.}(2014)\citenamefont
  {Bartoszek} \emph {et~al.}}]{Bartoszek:2014mya}%
  \BibitemOpen
  \bibfield  {author} {\bibinfo {author} {\bibfnamefont {L.}~\bibnamefont
  {Bartoszek}} \emph {et~al.} (\bibinfo {collaboration} {Mu2e}),\ }\href@noop
  {} {\  (\bibinfo {year} {2014})},\ \Eprint {http://arxiv.org/abs/1501.05241}
  {arXiv:1501.05241 [physics.ins-det]} \BibitemShut {NoStop}%
\bibitem [{\citenamefont {Porod}(2003)}]{Porod:2003um}%
  \BibitemOpen
  \bibfield  {author} {\bibinfo {author} {\bibfnamefont {W.}~\bibnamefont
  {Porod}},\ }\href {\doibase 10.1016/S0010-4655(03)00222-4} {\bibfield
  {journal} {\bibinfo  {journal} {Comput. Phys. Commun.}\ }\textbf {\bibinfo
  {volume} {153}},\ \bibinfo {pages} {275} (\bibinfo {year} {2003})},\ \Eprint
  {http://arxiv.org/abs/hep-ph/0301101} {arXiv:hep-ph/0301101 [hep-ph]}
  \BibitemShut {NoStop}%
\bibitem [{\citenamefont {Porod}\ and\ \citenamefont
  {Staub}(2012)}]{Porod:2011nf}%
  \BibitemOpen
  \bibfield  {author} {\bibinfo {author} {\bibfnamefont {W.}~\bibnamefont
  {Porod}}\ and\ \bibinfo {author} {\bibfnamefont {F.}~\bibnamefont {Staub}},\
  }\href {\doibase 10.1016/j.cpc.2012.05.021} {\bibfield  {journal} {\bibinfo
  {journal} {Comput. Phys. Commun.}\ }\textbf {\bibinfo {volume} {183}},\
  \bibinfo {pages} {2458} (\bibinfo {year} {2012})},\ \Eprint
  {http://arxiv.org/abs/1104.1573} {arXiv:1104.1573 [hep-ph]} \BibitemShut
  {NoStop}%
\bibitem [{\citenamefont {Porod}(2002)}]{Porod:2002wz}%
  \BibitemOpen
  \bibfield  {author} {\bibinfo {author} {\bibfnamefont {W.}~\bibnamefont
  {Porod}},\ }\href {\doibase 10.1088/1126-6708/2002/05/030} {\bibfield
  {journal} {\bibinfo  {journal} {JHEP}\ }\textbf {\bibinfo {volume} {05}},\
  \bibinfo {pages} {030} (\bibinfo {year} {2002})},\ \Eprint
  {http://arxiv.org/abs/hep-ph/0202259} {arXiv:hep-ph/0202259 [hep-ph]}
  \BibitemShut {NoStop}%
\bibitem [{\citenamefont {Staub}(2008)}]{Staub:2008uz}%
  \BibitemOpen
  \bibfield  {author} {\bibinfo {author} {\bibfnamefont {F.}~\bibnamefont
  {Staub}},\ }\href@noop {} {\  (\bibinfo {year} {2008})},\ \Eprint
  {http://arxiv.org/abs/0806.0538} {arXiv:0806.0538 [hep-ph]} \BibitemShut
  {NoStop}%
\bibitem [{\citenamefont {Staub}(2010)}]{Staub:2009bi}%
  \BibitemOpen
  \bibfield  {author} {\bibinfo {author} {\bibfnamefont {F.}~\bibnamefont
  {Staub}},\ }\href {\doibase 10.1016/j.cpc.2010.01.011} {\bibfield  {journal}
  {\bibinfo  {journal} {Comput. Phys. Commun.}\ }\textbf {\bibinfo {volume}
  {181}},\ \bibinfo {pages} {1077} (\bibinfo {year} {2010})},\ \Eprint
  {http://arxiv.org/abs/0909.2863} {arXiv:0909.2863 [hep-ph]} \BibitemShut
  {NoStop}%
\bibitem [{\citenamefont {Staub}(2011)}]{Staub:2010jh}%
  \BibitemOpen
  \bibfield  {author} {\bibinfo {author} {\bibfnamefont {F.}~\bibnamefont
  {Staub}},\ }\href {\doibase 10.1016/j.cpc.2010.11.030} {\bibfield  {journal}
  {\bibinfo  {journal} {Comput. Phys. Commun.}\ }\textbf {\bibinfo {volume}
  {182}},\ \bibinfo {pages} {808} (\bibinfo {year} {2011})},\ \Eprint
  {http://arxiv.org/abs/1002.0840} {arXiv:1002.0840 [hep-ph]} \BibitemShut
  {NoStop}%
\bibitem [{\citenamefont {Staub}(2014)}]{Staub:2013tta}%
  \BibitemOpen
  \bibfield  {author} {\bibinfo {author} {\bibfnamefont {F.}~\bibnamefont
  {Staub}},\ }\href {\doibase 10.1016/j.cpc.2014.02.018} {\bibfield  {journal}
  {\bibinfo  {journal} {Comput. Phys. Commun.}\ }\textbf {\bibinfo {volume}
  {185}},\ \bibinfo {pages} {1773} (\bibinfo {year} {2014})},\ \Eprint
  {http://arxiv.org/abs/1309.7223} {arXiv:1309.7223 [hep-ph]} \BibitemShut
  {NoStop}%
\bibitem [{\citenamefont {Staub}(2015)}]{Staub:2015kfa}%
  \BibitemOpen
  \bibfield  {author} {\bibinfo {author} {\bibfnamefont {F.}~\bibnamefont
  {Staub}},\ }\href {\doibase 10.1155/2015/840780} {\bibfield  {journal}
  {\bibinfo  {journal} {Adv. High Energy Phys.}\ }\textbf {\bibinfo {volume}
  {2015}},\ \bibinfo {pages} {840780} (\bibinfo {year} {2015})},\ \Eprint
  {http://arxiv.org/abs/1503.04200} {arXiv:1503.04200 [hep-ph]} \BibitemShut
  {NoStop}%
\bibitem [{\citenamefont {Adamson}\ \emph {et~al.}(2017)\citenamefont {Adamson}
  \emph {et~al.}}]{Adamson:2017gxd}%
  \BibitemOpen
  \bibfield  {author} {\bibinfo {author} {\bibfnamefont {P.}~\bibnamefont
  {Adamson}} \emph {et~al.} (\bibinfo {collaboration} {NOvA}),\ }\href
  {\doibase 10.1103/PhysRevLett.118.231801} {\bibfield  {journal} {\bibinfo
  {journal} {Phys. Rev. Lett.}\ }\textbf {\bibinfo {volume} {118}},\ \bibinfo
  {pages} {231801} (\bibinfo {year} {2017})},\ \Eprint
  {http://arxiv.org/abs/1703.03328} {arXiv:1703.03328 [hep-ex]} \BibitemShut
  {NoStop}%
\bibitem [{\citenamefont {Acero}\ \emph {et~al.}(2018)\citenamefont {Acero}
  \emph {et~al.}}]{NOvA:2018gge}%
  \BibitemOpen
  \bibfield  {author} {\bibinfo {author} {\bibfnamefont {M.~A.}\ \bibnamefont
  {Acero}} \emph {et~al.} (\bibinfo {collaboration} {NOvA}),\ }\href@noop {} {\
   (\bibinfo {year} {2018})},\ \Eprint {http://arxiv.org/abs/1806.00096}
  {arXiv:1806.00096 [hep-ex]} \BibitemShut {NoStop}%
\bibitem [{\citenamefont {Hinshaw}\ \emph {et~al.}(2013)\citenamefont {Hinshaw}
  \emph {et~al.}}]{Hinshaw:2012aka}%
  \BibitemOpen
  \bibfield  {author} {\bibinfo {author} {\bibfnamefont {G.}~\bibnamefont
  {Hinshaw}} \emph {et~al.} (\bibinfo {collaboration} {WMAP}),\ }\href
  {\doibase 10.1088/0067-0049/208/2/19} {\bibfield  {journal} {\bibinfo
  {journal} {Astrophys. J. Suppl.}\ }\textbf {\bibinfo {volume} {208}},\
  \bibinfo {pages} {19} (\bibinfo {year} {2013})},\ \Eprint
  {http://arxiv.org/abs/1212.5226} {arXiv:1212.5226 [astro-ph.CO]} \BibitemShut
  {NoStop}%
\bibitem [{\citenamefont {Aprile}\ \emph {et~al.}(2017)\citenamefont {Aprile}
  \emph {et~al.}}]{Aprile:2017iyp}%
  \BibitemOpen
  \bibfield  {author} {\bibinfo {author} {\bibfnamefont {E.}~\bibnamefont
  {Aprile}} \emph {et~al.} (\bibinfo {collaboration} {XENON}),\ }\href
  {\doibase 10.1103/PhysRevLett.119.181301} {\bibfield  {journal} {\bibinfo
  {journal} {Phys. Rev. Lett.}\ }\textbf {\bibinfo {volume} {119}},\ \bibinfo
  {pages} {181301} (\bibinfo {year} {2017})},\ \Eprint
  {http://arxiv.org/abs/1705.06655} {arXiv:1705.06655 [astro-ph.CO]}
  \BibitemShut {NoStop}%
\bibitem [{\citenamefont {Cui}\ \emph {et~al.}(2017)\citenamefont {Cui} \emph
  {et~al.}}]{Cui:2017nnn}%
  \BibitemOpen
  \bibfield  {author} {\bibinfo {author} {\bibfnamefont {X.}~\bibnamefont
  {Cui}} \emph {et~al.} (\bibinfo {collaboration} {PandaX-II}),\ }\href
  {\doibase 10.1103/PhysRevLett.119.181302} {\bibfield  {journal} {\bibinfo
  {journal} {Phys. Rev. Lett.}\ }\textbf {\bibinfo {volume} {119}},\ \bibinfo
  {pages} {181302} (\bibinfo {year} {2017})},\ \Eprint
  {http://arxiv.org/abs/1708.06917} {arXiv:1708.06917 [astro-ph.CO]}
  \BibitemShut {NoStop}%
\bibitem [{\citenamefont {Aprile}\ \emph {et~al.}(2018)\citenamefont {Aprile}
  \emph {et~al.}}]{Aprile:2018dbl}%
  \BibitemOpen
  \bibfield  {author} {\bibinfo {author} {\bibfnamefont {E.}~\bibnamefont
  {Aprile}} \emph {et~al.} (\bibinfo {collaboration} {XENON}),\ }\href@noop {}
  {\  (\bibinfo {year} {2018})},\ \Eprint {http://arxiv.org/abs/1805.12562}
  {arXiv:1805.12562 [astro-ph.CO]} \BibitemShut {NoStop}%
\bibitem [{\citenamefont {Arina}\ and\ \citenamefont
  {Cabrera}(2014)}]{Arina:2013zca}%
  \BibitemOpen
  \bibfield  {author} {\bibinfo {author} {\bibfnamefont {C.}~\bibnamefont
  {Arina}}\ and\ \bibinfo {author} {\bibfnamefont {M.~E.}\ \bibnamefont
  {Cabrera}},\ }\href {\doibase 10.1007/JHEP04(2014)100} {\bibfield  {journal}
  {\bibinfo  {journal} {JHEP}\ }\textbf {\bibinfo {volume} {04}},\ \bibinfo
  {pages} {100} (\bibinfo {year} {2014})},\ \Eprint
  {http://arxiv.org/abs/1311.6549} {arXiv:1311.6549 [hep-ph]} \BibitemShut
  {NoStop}%
\bibitem [{\citenamefont {Belanger}\ \emph {et~al.}(2014)\citenamefont
  {Belanger}, \citenamefont {Boudjema}, \citenamefont {Pukhov},\ and\
  \citenamefont {Semenov}}]{Belanger:2013oya}%
  \BibitemOpen
  \bibfield  {author} {\bibinfo {author} {\bibfnamefont {G.}~\bibnamefont
  {Belanger}}, \bibinfo {author} {\bibfnamefont {F.}~\bibnamefont {Boudjema}},
  \bibinfo {author} {\bibfnamefont {A.}~\bibnamefont {Pukhov}}, \ and\ \bibinfo
  {author} {\bibfnamefont {A.}~\bibnamefont {Semenov}},\ }\href {\doibase
  10.1016/j.cpc.2013.10.016} {\bibfield  {journal} {\bibinfo  {journal}
  {Comput. Phys. Commun.}\ }\textbf {\bibinfo {volume} {185}},\ \bibinfo
  {pages} {960} (\bibinfo {year} {2014})},\ \Eprint
  {http://arxiv.org/abs/1305.0237} {arXiv:1305.0237 [hep-ph]} \BibitemShut
  {NoStop}%
\bibitem [{\citenamefont {Aaboud}\ \emph
  {et~al.}(2018{\natexlab{b}})\citenamefont {Aaboud} \emph
  {et~al.}}]{Aaboud:2018jiw}%
  \BibitemOpen
  \bibfield  {author} {\bibinfo {author} {\bibfnamefont {M.}~\bibnamefont
  {Aaboud}} \emph {et~al.} (\bibinfo {collaboration} {ATLAS}),\ }\href@noop {}
  {\  (\bibinfo {year} {2018}{\natexlab{b}})},\ \Eprint
  {http://arxiv.org/abs/1803.02762} {arXiv:1803.02762 [hep-ex]} \BibitemShut
  {NoStop}%
\bibitem [{\citenamefont {Sirunyan}\ \emph {et~al.}(2018)\citenamefont
  {Sirunyan} \emph {et~al.}}]{Sirunyan:2018ubx}%
  \BibitemOpen
  \bibfield  {author} {\bibinfo {author} {\bibfnamefont {A.~M.}\ \bibnamefont
  {Sirunyan}} \emph {et~al.} (\bibinfo {collaboration} {CMS}),\ }\href@noop {}
  {\  (\bibinfo {year} {2018})},\ \Eprint {http://arxiv.org/abs/1801.03957}
  {arXiv:1801.03957 [hep-ex]} \BibitemShut {NoStop}%
\bibitem [{\citenamefont {Aaboud}\ \emph
  {et~al.}(2018{\natexlab{c}})\citenamefont {Aaboud} \emph
  {et~al.}}]{Aaboud:2017leg}%
  \BibitemOpen
  \bibfield  {author} {\bibinfo {author} {\bibfnamefont {M.}~\bibnamefont
  {Aaboud}} \emph {et~al.} (\bibinfo {collaboration} {ATLAS}),\ }\href
  {\doibase 10.1103/PhysRevD.97.052010} {\bibfield  {journal} {\bibinfo
  {journal} {Phys. Rev.}\ }\textbf {\bibinfo {volume} {D97}},\ \bibinfo {pages}
  {052010} (\bibinfo {year} {2018}{\natexlab{c}})},\ \Eprint
  {http://arxiv.org/abs/1712.08119} {arXiv:1712.08119 [hep-ex]} \BibitemShut
  {NoStop}%
\bibitem [{\citenamefont {Aaboud}\ \emph
  {et~al.}(2017{\natexlab{g}})\citenamefont {Aaboud} \emph
  {et~al.}}]{Aaboud:2017mpt}%
  \BibitemOpen
  \bibfield  {author} {\bibinfo {author} {\bibfnamefont {M.}~\bibnamefont
  {Aaboud}} \emph {et~al.} (\bibinfo {collaboration} {ATLAS}),\ }\href@noop {}
  {\  (\bibinfo {year} {2017}{\natexlab{g}})},\ \Eprint
  {http://arxiv.org/abs/1712.02118} {arXiv:1712.02118 [hep-ex]} \BibitemShut
  {NoStop}%
\bibitem [{\citenamefont {Sirunyan}\ \emph
  {et~al.}(2017{\natexlab{a}})\citenamefont {Sirunyan} \emph
  {et~al.}}]{Sirunyan:2017lae}%
  \BibitemOpen
  \bibfield  {author} {\bibinfo {author} {\bibfnamefont {A.~M.}\ \bibnamefont
  {Sirunyan}} \emph {et~al.} (\bibinfo {collaboration} {CMS}),\ }\href@noop {}
  {\  (\bibinfo {year} {2017}{\natexlab{a}})},\ \Eprint
  {http://arxiv.org/abs/1709.05406} {arXiv:1709.05406 [hep-ex]} \BibitemShut
  {NoStop}%
\bibitem [{\citenamefont {Aaboud}\ \emph
  {et~al.}(2018{\natexlab{d}})\citenamefont {Aaboud} \emph
  {et~al.}}]{Aaboud:2017nhr}%
  \BibitemOpen
  \bibfield  {author} {\bibinfo {author} {\bibfnamefont {M.}~\bibnamefont
  {Aaboud}} \emph {et~al.} (\bibinfo {collaboration} {ATLAS}),\ }\href
  {\doibase 10.1140/epjc/s10052-018-5583-9} {\bibfield  {journal} {\bibinfo
  {journal} {Eur. Phys. J.}\ }\textbf {\bibinfo {volume} {C78}},\ \bibinfo
  {pages} {154} (\bibinfo {year} {2018}{\natexlab{d}})},\ \Eprint
  {http://arxiv.org/abs/1708.07875} {arXiv:1708.07875 [hep-ex]} \BibitemShut
  {NoStop}%
\bibitem [{\citenamefont {Sirunyan}\ \emph
  {et~al.}(2017{\natexlab{b}})\citenamefont {Sirunyan} \emph
  {et~al.}}]{Sirunyan:2017zss}%
  \BibitemOpen
  \bibfield  {author} {\bibinfo {author} {\bibfnamefont {A.~M.}\ \bibnamefont
  {Sirunyan}} \emph {et~al.} (\bibinfo {collaboration} {CMS}),\ }\href
  {\doibase 10.1007/JHEP11(2017)029} {\bibfield  {journal} {\bibinfo  {journal}
  {JHEP}\ }\textbf {\bibinfo {volume} {11}},\ \bibinfo {pages} {029} (\bibinfo
  {year} {2017}{\natexlab{b}})},\ \Eprint {http://arxiv.org/abs/1706.09933}
  {arXiv:1706.09933 [hep-ex]} \BibitemShut {NoStop}%
\bibitem [{\citenamefont {Drees}\ \emph {et~al.}(2015)\citenamefont {Drees},
  \citenamefont {Dreiner}, \citenamefont {Schmeier}, \citenamefont
  {Tattersall},\ and\ \citenamefont {Kim}}]{Drees:2013wra}%
  \BibitemOpen
  \bibfield  {author} {\bibinfo {author} {\bibfnamefont {M.}~\bibnamefont
  {Drees}}, \bibinfo {author} {\bibfnamefont {H.}~\bibnamefont {Dreiner}},
  \bibinfo {author} {\bibfnamefont {D.}~\bibnamefont {Schmeier}}, \bibinfo
  {author} {\bibfnamefont {J.}~\bibnamefont {Tattersall}}, \ and\ \bibinfo
  {author} {\bibfnamefont {J.~S.}\ \bibnamefont {Kim}},\ }\href {\doibase
  10.1016/j.cpc.2014.10.018} {\bibfield  {journal} {\bibinfo  {journal}
  {Comput. Phys. Commun.}\ }\textbf {\bibinfo {volume} {187}},\ \bibinfo
  {pages} {227} (\bibinfo {year} {2015})},\ \Eprint
  {http://arxiv.org/abs/1312.2591} {arXiv:1312.2591 [hep-ph]} \BibitemShut
  {NoStop}%
\bibitem [{\citenamefont {Dercks}\ \emph {et~al.}(2017)\citenamefont {Dercks},
  \citenamefont {Desai}, \citenamefont {Kim}, \citenamefont {Rolbiecki},
  \citenamefont {Tattersall},\ and\ \citenamefont {Weber}}]{Dercks:2016npn}%
  \BibitemOpen
  \bibfield  {author} {\bibinfo {author} {\bibfnamefont {D.}~\bibnamefont
  {Dercks}}, \bibinfo {author} {\bibfnamefont {N.}~\bibnamefont {Desai}},
  \bibinfo {author} {\bibfnamefont {J.~S.}\ \bibnamefont {Kim}}, \bibinfo
  {author} {\bibfnamefont {K.}~\bibnamefont {Rolbiecki}}, \bibinfo {author}
  {\bibfnamefont {J.}~\bibnamefont {Tattersall}}, \ and\ \bibinfo {author}
  {\bibfnamefont {T.}~\bibnamefont {Weber}},\ }\href {\doibase
  10.1016/j.cpc.2017.08.021} {\bibfield  {journal} {\bibinfo  {journal}
  {Comput. Phys. Commun.}\ }\textbf {\bibinfo {volume} {221}},\ \bibinfo
  {pages} {383} (\bibinfo {year} {2017})},\ \Eprint
  {http://arxiv.org/abs/1611.09856} {arXiv:1611.09856 [hep-ph]} \BibitemShut
  {NoStop}%
\bibitem [{\citenamefont {Alwall}\ \emph {et~al.}(2011)\citenamefont {Alwall},
  \citenamefont {Herquet}, \citenamefont {Maltoni}, \citenamefont {Mattelaer},\
  and\ \citenamefont {Stelzer}}]{Alwall:2011uj}%
  \BibitemOpen
  \bibfield  {author} {\bibinfo {author} {\bibfnamefont {J.}~\bibnamefont
  {Alwall}}, \bibinfo {author} {\bibfnamefont {M.}~\bibnamefont {Herquet}},
  \bibinfo {author} {\bibfnamefont {F.}~\bibnamefont {Maltoni}}, \bibinfo
  {author} {\bibfnamefont {O.}~\bibnamefont {Mattelaer}}, \ and\ \bibinfo
  {author} {\bibfnamefont {T.}~\bibnamefont {Stelzer}},\ }\href {\doibase
  10.1007/JHEP06(2011)128} {\bibfield  {journal} {\bibinfo  {journal} {JHEP}\
  }\textbf {\bibinfo {volume} {06}},\ \bibinfo {pages} {128} (\bibinfo {year}
  {2011})},\ \Eprint {http://arxiv.org/abs/1106.0522} {arXiv:1106.0522
  [hep-ph]} \BibitemShut {NoStop}%
\bibitem [{\citenamefont {Alwall}\ \emph {et~al.}(2014)\citenamefont {Alwall},
  \citenamefont {Frederix}, \citenamefont {Frixione}, \citenamefont {Hirschi},
  \citenamefont {Maltoni}, \citenamefont {Mattelaer}, \citenamefont {Shao},
  \citenamefont {Stelzer}, \citenamefont {Torrielli},\ and\ \citenamefont
  {Zaro}}]{Alwall:2014hca}%
  \BibitemOpen
  \bibfield  {author} {\bibinfo {author} {\bibfnamefont {J.}~\bibnamefont
  {Alwall}}, \bibinfo {author} {\bibfnamefont {R.}~\bibnamefont {Frederix}},
  \bibinfo {author} {\bibfnamefont {S.}~\bibnamefont {Frixione}}, \bibinfo
  {author} {\bibfnamefont {V.}~\bibnamefont {Hirschi}}, \bibinfo {author}
  {\bibfnamefont {F.}~\bibnamefont {Maltoni}}, \bibinfo {author} {\bibfnamefont
  {O.}~\bibnamefont {Mattelaer}}, \bibinfo {author} {\bibfnamefont {H.~S.}\
  \bibnamefont {Shao}}, \bibinfo {author} {\bibfnamefont {T.}~\bibnamefont
  {Stelzer}}, \bibinfo {author} {\bibfnamefont {P.}~\bibnamefont {Torrielli}},
  \ and\ \bibinfo {author} {\bibfnamefont {M.}~\bibnamefont {Zaro}},\ }\href
  {\doibase 10.1007/JHEP07(2014)079} {\bibfield  {journal} {\bibinfo  {journal}
  {JHEP}\ }\textbf {\bibinfo {volume} {07}},\ \bibinfo {pages} {079} (\bibinfo
  {year} {2014})},\ \Eprint {http://arxiv.org/abs/1405.0301} {arXiv:1405.0301
  [hep-ph]} \BibitemShut {NoStop}%
\bibitem [{\citenamefont {Sjöstrand}\ \emph {et~al.}(2015)\citenamefont
  {Sjöstrand}, \citenamefont {Ask}, \citenamefont {Christiansen},
  \citenamefont {Corke}, \citenamefont {Desai}, \citenamefont {Ilten},
  \citenamefont {Mrenna}, \citenamefont {Prestel}, \citenamefont {Rasmussen},\
  and\ \citenamefont {Skands}}]{Sjostrand:2014zea}%
  \BibitemOpen
  \bibfield  {author} {\bibinfo {author} {\bibfnamefont {T.}~\bibnamefont
  {Sjöstrand}}, \bibinfo {author} {\bibfnamefont {S.}~\bibnamefont {Ask}},
  \bibinfo {author} {\bibfnamefont {J.~R.}\ \bibnamefont {Christiansen}},
  \bibinfo {author} {\bibfnamefont {R.}~\bibnamefont {Corke}}, \bibinfo
  {author} {\bibfnamefont {N.}~\bibnamefont {Desai}}, \bibinfo {author}
  {\bibfnamefont {P.}~\bibnamefont {Ilten}}, \bibinfo {author} {\bibfnamefont
  {S.}~\bibnamefont {Mrenna}}, \bibinfo {author} {\bibfnamefont
  {S.}~\bibnamefont {Prestel}}, \bibinfo {author} {\bibfnamefont {C.~O.}\
  \bibnamefont {Rasmussen}}, \ and\ \bibinfo {author} {\bibfnamefont {P.~Z.}\
  \bibnamefont {Skands}},\ }\href {\doibase 10.1016/j.cpc.2015.01.024}
  {\bibfield  {journal} {\bibinfo  {journal} {Comput. Phys. Commun.}\ }\textbf
  {\bibinfo {volume} {191}},\ \bibinfo {pages} {159} (\bibinfo {year}
  {2015})},\ \Eprint {http://arxiv.org/abs/1410.3012} {arXiv:1410.3012
  [hep-ph]} \BibitemShut {NoStop}%
\bibitem [{\citenamefont {Ball}\ \emph {et~al.}(2013)\citenamefont {Ball} \emph
  {et~al.}}]{Ball:2012cx}%
  \BibitemOpen
  \bibfield  {author} {\bibinfo {author} {\bibfnamefont {R.~D.}\ \bibnamefont
  {Ball}} \emph {et~al.},\ }\href {\doibase 10.1016/j.nuclphysb.2012.10.003}
  {\bibfield  {journal} {\bibinfo  {journal} {Nucl. Phys.}\ }\textbf {\bibinfo
  {volume} {B867}},\ \bibinfo {pages} {244} (\bibinfo {year} {2013})},\ \Eprint
  {http://arxiv.org/abs/1207.1303} {arXiv:1207.1303 [hep-ph]} \BibitemShut
  {NoStop}%
\bibitem [{\citenamefont {Ball}\ \emph {et~al.}(2015)\citenamefont {Ball} \emph
  {et~al.}}]{Ball:2014uwa}%
  \BibitemOpen
  \bibfield  {author} {\bibinfo {author} {\bibfnamefont {R.~D.}\ \bibnamefont
  {Ball}} \emph {et~al.} (\bibinfo {collaboration} {NNPDF}),\ }\href {\doibase
  10.1007/JHEP04(2015)040} {\bibfield  {journal} {\bibinfo  {journal} {JHEP}\
  }\textbf {\bibinfo {volume} {04}},\ \bibinfo {pages} {040} (\bibinfo {year}
  {2015})},\ \Eprint {http://arxiv.org/abs/1410.8849} {arXiv:1410.8849
  [hep-ph]} \BibitemShut {NoStop}%
\bibitem [{\citenamefont {de~Favereau}\ \emph {et~al.}(2014)\citenamefont
  {de~Favereau}, \citenamefont {Delaere}, \citenamefont {Demin}, \citenamefont
  {Giammanco}, \citenamefont {Lemaître}, \citenamefont {Mertens},\ and\
  \citenamefont {Selvaggi}}]{deFavereau:2013fsa}%
  \BibitemOpen
  \bibfield  {author} {\bibinfo {author} {\bibfnamefont {J.}~\bibnamefont
  {de~Favereau}}, \bibinfo {author} {\bibfnamefont {C.}~\bibnamefont
  {Delaere}}, \bibinfo {author} {\bibfnamefont {P.}~\bibnamefont {Demin}},
  \bibinfo {author} {\bibfnamefont {A.}~\bibnamefont {Giammanco}}, \bibinfo
  {author} {\bibfnamefont {V.}~\bibnamefont {Lemaître}}, \bibinfo {author}
  {\bibfnamefont {A.}~\bibnamefont {Mertens}}, \ and\ \bibinfo {author}
  {\bibfnamefont {M.}~\bibnamefont {Selvaggi}} (\bibinfo {collaboration}
  {DELPHES 3}),\ }\href {\doibase 10.1007/JHEP02(2014)057} {\bibfield
  {journal} {\bibinfo  {journal} {JHEP}\ }\textbf {\bibinfo {volume} {02}},\
  \bibinfo {pages} {057} (\bibinfo {year} {2014})},\ \Eprint
  {http://arxiv.org/abs/1307.6346} {arXiv:1307.6346 [hep-ex]} \BibitemShut
  {NoStop}%
\bibitem [{\citenamefont {Selvaggi}(2014)}]{Selvaggi:2014mya}%
  \BibitemOpen
  \bibfield  {author} {\bibinfo {author} {\bibfnamefont {M.}~\bibnamefont
  {Selvaggi}},\ }\bibfield  {booktitle} {\emph {\bibinfo {booktitle}
  {{Proceedings, 15th International Workshop on Advanced Computing and Analysis
  Techniques in Physics Research (ACAT 2013)}}},\ }\href {\doibase
  10.1088/1742-6596/523/1/012033} {\bibfield  {journal} {\bibinfo  {journal}
  {J. Phys. Conf. Ser.}\ }\textbf {\bibinfo {volume} {523}},\ \bibinfo {pages}
  {012033} (\bibinfo {year} {2014})}\BibitemShut {NoStop}%
\bibitem [{\citenamefont {Mertens}(2015)}]{Mertens:2015kba}%
  \BibitemOpen
  \bibfield  {author} {\bibinfo {author} {\bibfnamefont {A.}~\bibnamefont
  {Mertens}},\ }\bibfield  {booktitle} {\emph {\bibinfo {booktitle}
  {{Proceedings, 16th International workshop on Advanced Computing and Analysis
  Techniques in physics (ACAT 14)}}},\ }\href {\doibase
  10.1088/1742-6596/608/1/012045} {\bibfield  {journal} {\bibinfo  {journal}
  {J. Phys. Conf. Ser.}\ }\textbf {\bibinfo {volume} {608}},\ \bibinfo {pages}
  {012045} (\bibinfo {year} {2015})}\BibitemShut {NoStop}%
\bibitem [{\citenamefont {Cacciari}\ \emph {et~al.}(2008)\citenamefont
  {Cacciari}, \citenamefont {Salam},\ and\ \citenamefont
  {Soyez}}]{Cacciari:2008gp}%
  \BibitemOpen
  \bibfield  {author} {\bibinfo {author} {\bibfnamefont {M.}~\bibnamefont
  {Cacciari}}, \bibinfo {author} {\bibfnamefont {G.~P.}\ \bibnamefont {Salam}},
  \ and\ \bibinfo {author} {\bibfnamefont {G.}~\bibnamefont {Soyez}},\ }\href
  {\doibase 10.1088/1126-6708/2008/04/063} {\bibfield  {journal} {\bibinfo
  {journal} {JHEP}\ }\textbf {\bibinfo {volume} {04}},\ \bibinfo {pages} {063}
  (\bibinfo {year} {2008})},\ \Eprint {http://arxiv.org/abs/0802.1189}
  {arXiv:0802.1189 [hep-ph]} \BibitemShut {NoStop}%
\bibitem [{cse()}]{csec:stop}%
  \BibitemOpen
  \href {\doibase
  https://twiki.cern.ch/twiki/bin/view/LHCPhysics/SUSYCrossSections} {\
  https://twiki.cern.ch/twiki/bin/view/LHCPhysics/SUSYCrossSections}\BibitemShut
  {NoStop}%
\bibitem [{\citenamefont {Beenakker}\ \emph {et~al.}(1996)\citenamefont
  {Beenakker}, \citenamefont {Hopker},\ and\ \citenamefont
  {Spira}}]{Beenakker:1996ed}%
  \BibitemOpen
  \bibfield  {author} {\bibinfo {author} {\bibfnamefont {W.}~\bibnamefont
  {Beenakker}}, \bibinfo {author} {\bibfnamefont {R.}~\bibnamefont {Hopker}}, \
  and\ \bibinfo {author} {\bibfnamefont {M.}~\bibnamefont {Spira}},\
  }\href@noop {} {\  (\bibinfo {year} {1996})},\ \Eprint
  {http://arxiv.org/abs/hep-ph/9611232} {arXiv:hep-ph/9611232 [hep-ph]}
  \BibitemShut {NoStop}%
\bibitem [{\citenamefont {Beenakker}\ \emph {et~al.}(1998)\citenamefont
  {Beenakker}, \citenamefont {Kramer}, \citenamefont {Plehn}, \citenamefont
  {Spira},\ and\ \citenamefont {Zerwas}}]{Beenakker:1997ut}%
  \BibitemOpen
  \bibfield  {author} {\bibinfo {author} {\bibfnamefont {W.}~\bibnamefont
  {Beenakker}}, \bibinfo {author} {\bibfnamefont {M.}~\bibnamefont {Kramer}},
  \bibinfo {author} {\bibfnamefont {T.}~\bibnamefont {Plehn}}, \bibinfo
  {author} {\bibfnamefont {M.}~\bibnamefont {Spira}}, \ and\ \bibinfo {author}
  {\bibfnamefont {P.~M.}\ \bibnamefont {Zerwas}},\ }\href {\doibase
  10.1016/S0550-3213(98)00014-5} {\bibfield  {journal} {\bibinfo  {journal}
  {Nucl. Phys.}\ }\textbf {\bibinfo {volume} {B515}},\ \bibinfo {pages} {3}
  (\bibinfo {year} {1998})},\ \Eprint {http://arxiv.org/abs/hep-ph/9710451}
  {arXiv:hep-ph/9710451 [hep-ph]} \BibitemShut {NoStop}%
\end{thebibliography}%
\end{document}